\documentclass[aoas,MSNbibl,nameyear,seceqn,dvips]{arximspdf}
\usepackage{dcolumn}
\usepackage{graphicx}
\doi{10.1214/11-AOAS459}
\volume{5}
\issue{3}
\pubyear{2011}
\firstpage{2078}
\lastpage{2108}

\makeatletter
\def\bptnote#1{}
\def\bsuffix #1{#1}
\newcolumntype{d}[1]{D{.}{.}{#1}}
\newcommand{\cal}{\mathcal}
\def\bs{\mathbf}
\makeatother

\begin{document}
\begin{frontmatter}

\title{Semiparametric modeling of autonomous nonlinear dynamical
systems with application to plant~growth}
\runtitle{Modeling autonomous nonlinear dynamical systems}

\begin{aug}
\author[A]{\fnms{Debashis} \snm{Paul}\corref{}\thanksref{t1}\ead[label=e1]{debashis@wald.ucdavis.edu}},
\author[A]{\fnms{Jie} \snm{Peng}\thanksref{t2}\ead[label=e2]{jie@wald.ucdavis.edu}}
\and
\author[A]{\fnms{Prabir} \snm{Burman}\thanksref{t3}\ead[label=e3]{burman@wald.ucdavis.edu}}
\runauthor{D. Paul, J. Peng and P. Burman}
\affiliation{University of California, Davis}
\address[A]{Department of Statistics\\
University of California\\
Davis, California 95616\\USA\\
\printead{e1}\\
\phantom{\textsc{E-mail}: }\printead*{e2} \\
\phantom{\textsc{E-mail}: }\printead*{e3}}
\end{aug}
\thankstext{t1}{Supported by the NSF Grants DMS-08-06128
and DMR-10-35468.}
\thankstext{t2}{Supported by the NSF Grants DMS-08-06128 and DMS-10-01256.}
\thankstext{t3}{Supported by the NSF Grant
DMS-09-07622.}

\received{\smonth{5} \syear{2010}}
\revised{\smonth{12} \syear{2010}}

%
\begin{abstract}
We propose a semiparametric model for autonomous nonlinear
dynamical systems and devise an estimation procedure for model
fitting. This model incorporates subject-specific effects and can be
viewed as a nonlinear semiparametric mixed effects model. We also
propose a computationally efficient model selection procedure. We
show by simulation studies that the proposed estimation as well as
model selection procedures can efficiently handle sparse and noisy
measurements. Finally, we apply the proposed method to a plant
growth data used to study growth displacement rates within meristems
of maize roots under two different experimental conditions.
\end{abstract}

%

\end{frontmatter}

\section{Introduction}\label{sec:intro}\label{sec1}

Continuous time dynamical systems arise, among other places, in
modeling certain biological processes. This includes classical
examples from population biology like the \textit{Lotka--Voltera
equations} for describing prey-predator dynamics
[\citet{Perthame2007}], or subject-specific processes like the
progression of infectious diseases in individuals
[\citet{NowakMay2000}]. Most of the existing
approaches estimate the dynamical system by assuming known
functional forms of the system. Moreover, many of them aim at
estimating individual dynamics for one subject. However, in many
scientific studies, there is a need to model the dynamical system
nonparametrically due to insufficient knowledge of the problem at
hand. In addition, there could be an interest to know the
dynamics of a certain process at a population level in order to
answer various scientific questions. Thus, in this paper, we
propose a new method to bridge the gap and tackle these
challenges.\looseness=1

%
\begin{figure}

\includegraphics{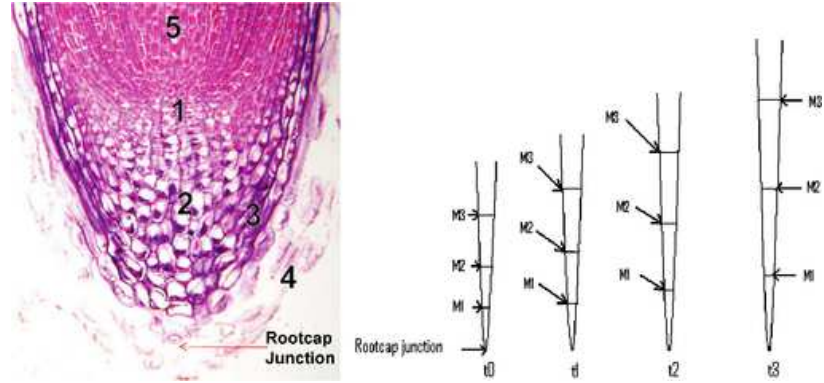}

\caption{Root tip. Left panel: image of root tip with meristem$^*$: 1---meristem;
4---root cap; 5---elongation zone. Right panel: an illustration of the
root tip
with the displacements of three markers $M1$, $M2$, $M3$ indicated at times
$t_0, t_1, t_2, t_3$. ($^*$From wikipedia.)}
\label{figure:meristem}
\end{figure}

To motivate the model, we first briefly discuss a study on plant
growth. There is a lot of research aiming to understand the effect
of environmental conditions on the growth in a plant. For example,
root growth in plants is highly sensitive to environmental factors
such as temperature, water deficit or nutrients
[\citet{Schurr2006}; \citet{Walter2002}]. In \citet{Sacks1997}, an experiment
is conducted to study the effect of water stress on cortical cell
division rates through \textit{growth displacement rate} within the
meristem of the primary root of maize seedlings
(Figure~\ref{figure:meristem}: left panel). In this study, for each plant,
measurements are taken on the displacement, measured as the distance
in millimeters from the root cap junction (root apex) of a number
of markers on the root over a period of $12$ hours (Figure
\ref{figure:meristem}: right panel). The growth displacement rate is
defined as the rate of displacement of a particle placed along the
root and, thus, it is a function of distance from the root apex. By
its definition, growth displacement rate characterizes the
relationship between the growth trajectory and its derivative (with
respect to time). Therefore, it is the gradient function in the
corresponding dynamical system. In this study there is a~need to
understand the dynamics at the population level, while accounting
for subject-specific variations, in order to compare the growth
displacement rates under two different water conditions.

Motivated by this study, in this paper, we focus on modeling and
fitting the underlying dynamical system based on data measured over
time, referred to as sample curves or sample paths, for a group of
subjects. Moreover, for a given sample curve, instead of observing
the whole sample path, measurements are taken only at a sparse set
of time points together with possible measurement noise. In the
plant data that we just mentioned, each plant is a subject, and the
positions of the markers which are located at different distances at
time zero from the root cap junction correspond to different initial
conditions. Each marker corresponds to one displacement trajectory
(also referred to as growth trajectory/curve), and the number of
measurements varies from two to seventeen, with measurement times
varying across trajectories. (See Section \ref{sec:plant} for a
more detailed description.)

We first give a brief overview of the existing literature on fitting
smooth deterministic dynamical systems in continuous time. A large
number of physical, chemical or biological processes are modeled
through systems of parametric differential equations
[\citet{LjungGlad1994}; \citet{Perthame2007}; \citet{Strogatz2001}]. For example,
\citet{Ramsay2007} consider modeling a~con\-tinuously stirred tank
reactor and propose a method called parameter cascading for model
fitting. \citet{ZhuWu2007} adopt a state space approach for
estimating the dynamics of cell-virus interactions in an AIDS
clinical trial. \citeauthor{RamsaySilverman2002} (\citeyear{RamsaySilverman2002,RamsaySilverman2005})
consider fitting dynamical systems given by systems of linear
differential equations where the coefficients of the differential
operator may be time varying. They propose methods for estimating
these (linear) differential operators based on principal
differential analysis when the data are recorded at dense and
regular time points. \citet{Poyton2006} also use the principal
differential analysis approach to fit dynamical systems.
\citeauthor{ChenWu2008a} (\citeyear{ChenWu2008a,ChenWu2008b}) propose to estimate parametric
differential equations with known functional forms and
time-dependent parameters through a two-stage approach where the
first stage involves estimation of the sample trajectories and their
derivatives by nonparametric smoothing. \citet{Brunel2008} gives a
comprehensive theoretical analysis of such an approach.
\citet{Cao2008} propose a~method for fitting nonlinear dynamical
systems using splines with predetermined knots for describing the
gradient function. This involves knowing the functional form of the
differential equation and does not include any subject-specific
effects. \citet{WuDing1999} and \citet{Wu1998} propose using
the nonlinear least squares procedure for fitting parametric
differential equations that take into account subject-specific
effects.

For the problems that we address in this paper, measurements are
taken on a sparse set of points for each sample curve so that
estimation of individual sample trajectory or its derivative based
on nonparametric smoothing is error-prone and results in a loss of
information. Thus, numerical procedures for solving differential
equations can become unstable if we treat each sample curve
separately. Moreover, we are more interested in estimating the
baseline dynamics at the population level than the individual
dynamics of each subject. For example, in the plant study described
above, we are interested in comparing the growth displacement rates
under two different experimental conditions. On the other hand, we
are not so interested in the individual displacement rate
corresponding to each plant. Another important aspect in modeling
data with multiple subjects is that adequate measures need to be
taken to model possible subject-specific effects, otherwise the
estimates of the model parameters can have inflated variability. In
this paper we propose a semiparametric approach for modeling
dynamical systems which incorporates subject-specific effects while
combining information across different subjects. A nonparametric
model is often essential because of insufficient knowledge about the
problem to suggest a reasonable parametric form of the dynamical
system. In addition, if realistic parametric models can be proposed,
then the nonparametric fit can be used for diagnostics of lack of
fit, for example, by employing a distance measure between the
parametric and nonparametric fits and studying its sampling
variability. We propose an estimation procedure that combines
nonlinear optimization techniques with a numerical ODE (ordinary
differential equation) solver to estimate the unknown parameters. In
addition, we derive a computationally efficient approximation of the
leave-one-curve-out cross-validation score for model selection. We
show by simulation studies that the proposed approach can
efficiently estimate the baseline dynamics with noisy and sparsely
measured sample curves. Finally, we apply the proposed method to the
plant data and compare the estimated growth displacement rates under
the two experimental conditions and discuss some scientific
implications of the results.

To the best of our knowledge, modeling and fitting dynamical
systems nonparametrically while also allowing for subject-specific
effects is new in the literature. In particular, our model differs
from traditional nonlinear mixed effects models previously employed
for fitting differential equations, which are almost exclusively
parametric [\citet{Wu1998}; \citet{WuDing1999}; \citet{Guedj2007}; \citet{Li2002}]. For
example, \citet{Guedj2007} consider a~nonlinear state-space model
where the state variable follows a parametric differential equation
with subject-specific effects, and the parameters are estimated
through a maximum likelihood approach. In contrast, for the model
proposed in this paper, the form of the gradient function $g$ is not
assumed to be known and it is approximated in a sequence of bases
with growing dimension. Note that this gives rise to a sequence of
parametric models with increasing complexity, and one needs to
adopt a model selection procedure to select an appropriate model, as
is typical in nonparametric function estimation. The theoretical
derivations in \citet{Paul2009} also show that the problem of
estimating the gradient function $g$ nonparametrically is
intrinsically different from that under a parametric nonlinear
mixed-effects model.

The rest of the paper is organized as follows. In Section
\ref{sec:model} we describe the proposed model. In Section
\ref{sec:estimation} we discuss the model fitting and model
selection procedures. In Section \ref{sec:simulation} we conduct
simulation studies to illustrate finite sample performance of the
proposed method and compare the proposed method with a two-stage
procedure. In Section \ref{sec:plant} we apply this method to the
plant data. Section~\ref{sec:dis} has a brief discussion. More
details and additional simulation results are reported in the
supplementary material [\citet{Paul-supp2011}].

\section{Model}\label{sec:model}\label{sec2}

In this section we describe a class of autonomous dynamical systems
that is suitable for modeling the problems discussed in Section \ref
{sec:intro}.
An autonomous dynamical system has the following general form:
\[
X^{\prime}(t) = f(X(t)),  \qquad  t \in[T_0,T_1].
\]
Without loss of generality, henceforth $T_0=0$ and $T_1 = 1$. Note
that the above equation implies that $X(t) = a + \int_0^t f(X(u))\,du$, where $a = X(0)$ is the initial condition. In an autonomous
system, the dynamics, which is characterized by $f$, depends on time
$t$ only through the ``state'' $X(t)$. This type of system arises in
various scientific studies such as modeling prey-predator dynamics,
virus dynamics or epidemiology [\citet{Perthame2007}].

In this paper we consider the following class of
autonomous dynamical systems:
%
\begin{equation}\label{eq:basic}
X_{il}'(t) = g_i(X_{il}(t)), \qquad  l=1,\ldots , N_i, \ i=1,\ldots ,n,
\end{equation}
where $\{X_{il}(t)\dvtx t \in[0,1], l=1,\ldots , N_i; i=1,\ldots ,n\}$ is
a collection of smooth curves corresponding to $n$ subjects, where
$N_i \geq1$ is the number of curves associated with the $i$th
subject. For example, in the plant study, each plant is a subject
and each marker corresponds to one growth curve and there are
multiple markers for each plant. We assume that all the curves
associated with the same subject follow the same dynamics and are
only differentiated by different initial conditions. These are
described by the functions $\{g_i(\cdot)\}_{i=1}^n$. In this paper
we model $\{g_i(\cdot)\}_{i=1}^n$ as
%
\begin{equation}\label{eq:scale}
g_i(\cdot) = e^{\theta_i} g(\cdot), \qquad i=1,\ldots ,n,
\end{equation}
where:
\begin{longlist}[(2)]
\item[(1)] the function $g(\cdot)$ reflects the common
underlying mechanism regulating all these dynamical systems. It is
assumed to be a smooth function and is referred to as the \textit{gradient function}.

\item[(2)] $\theta_i$'s reflect subject-specific effects in these systems.
The mean of $\theta_i$'s is assumed to be zero to impose
identifiability.
\end{longlist}
Note that one may view the trajectories for each plant as
multivariate functional data. However, here for each subject, the
different trajectories correspond to different initial conditions of
the same ODE describing the system. This means that given the
initial condition and the subject-specific scaling parameter
$\theta_i$, the corresponding trajectory is completely determined by
the underlying dynamical system and the only source of randomness is
from measurement errors.

\begin{figure}

\includegraphics{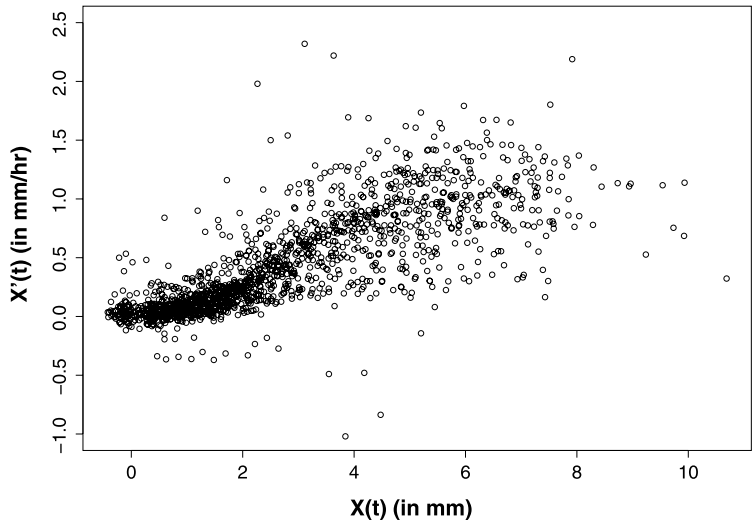}

\caption{Empirical derivatives $\widehat X'(t)$ against empirical
fits $\widehat X(t)$ for the treatment group in the plant growth
data.} \label{figure:emperical_scatter}
\end{figure}

The simplicity and generality of this model make it appealing for
modeling a wide class of dynamical systems. First, the gradient
function $g(\cdot)$ can be an arbitrary smooth function. Second,
the scale parameter $e^{\theta_i}$ provides a~subject-specific
tuning of the dynamics. This is motivated by the fact that, for a
large class of problems, the variations of the dynamics in a
population are in the scale of the rate of change rather than in the
shape of the gradient function. For example, for the plant data, by
examining the scatter plot of empirical derivatives versus
empirical fits (Figure \ref{figure:emperical_scatter}, for more
details, see Section~\ref{sec5}), we observe an excessive variability toward
the end which reflects plant-specific scaling effects. Moreover, the
above model is also flexible in incorporating time-independent
covariates, say, $z_i$, for example, by expressing the scaling factor
as $e^{\eta^T z_i}$ for some parameter $\eta$. In this paper our
primary goal is to estimate the gradient function $g$
nonparametrically.

%

Assuming the gradient function $g$ to be smooth means that it can be
well-approximated by a basis representation approach:
\[\label{eq:basis}
g(x) \approx\sum_{k=1}^M \beta_k \phi_{k,M}(x),
\]
where $\phi_{1,M}(\cdot),\ldots ,\phi_{M,M}(\cdot)$ are linearly
independent basis functions, chosen so that their combined support
covers the range of the observed trajectories. For example, we can
use cubic splines with a suitable set of knots. Thus, for a~given
choice of the basis functions, the unknown parameters in the model
are the basis coefficients $\bolds{\beta}
:=(\beta_1,\ldots ,\beta_M)^T$, the scale parameters $\bolds{\theta}
:=\{\theta_i\}_{i=1}^n$, and possibly the initial conditions $\bs{a}
:= \{a_{il} := X_{il}(0)\dvtx  l=1,\ldots ,N_i\}_{i=1}^n$. Also, various
model parameters, such as the number of basis functions $M$ and the
knot sequence, need to be selected based on the data. Therefore, in
essence, this is a nonlinear, semiparametric, mixed effects model.

\section{Model fitting}\label{sec:estimation}\label{sec3}

\subsection{Estimation procedure}\label{subsec:modelfit}\label{sec3.1}

In this section we propose an estimation procedure based on
sparsely observed noisy data. Specifically, we assume that the
observations are given by
%
\begin{equation}\label{eq:data_model}
Y_{ilj} = X_{il}(t_{ilj}) + \varepsilon_{ilj}, \qquad j=1,\ldots ,m_{il},
\end{equation}
where $0 \leq t_{il1} < \cdots< t_{ilm_{il}} \leq1$ are the
measurement times for the $l$th curve of the $i$th subject, and
$\{\varepsilon_{ilj}\}$ are independently and identically
distributed noise with mean zero and variance
$\sigma^2_{\varepsilon} > 0$. For model fitting with such data, we
adopt an iterative updating procedure which imposes regularization
on the estimates of $\bolds{\theta}$ and~$\bs{a}$. One way to achieve
this is to treat them as unknown random parameters from some
parametric distributions. Specifically, we use the following set of
working assumptions: (i) $a_{il}$'s are independent and identically
distributed as $N(\alpha,\sigma_a^2)$ and $\theta_i$'s are
independent and identically distributed as $N(0,\sigma_\theta^2)$,
for some $\alpha\in\mathbb{R}$ and $\sigma_a^2 > 0,
\sigma_\theta^2>0$; (ii) the noise $\varepsilon_{ilj}$'s are
independent and identically distributed as
$N(0,\sigma_\varepsilon^2)$ for $\sigma_\varepsilon^2
> 0$; (iii) the three random vectors $\bs{a},$ $\bolds{\theta}$,
$\bolds{\varepsilon} := \{\varepsilon_{ilj}\}$ are independent. Under
these assumptions, the negative joint log-likelihood of the observed
data $Y :=\{Y_{ilj}\}$, the scale parameters $\bolds{\theta}$ and the
initial conditions $\bs{a}$ is (up to an additive constant and a
positive scale constant)
%
\begin{equation}\label{eq:log-like}
 \qquad \sum_{i=1}^n \sum_{l=1}^{N_i}\sum_{j=1}^{m_{il}}
[Y_{ilj} - \widetilde X_{il}(t_{ilj};a_{il},\theta_i,\bolds\beta)]^2
+ \lambda_1
\sum_{i=1}^n\sum_{l=1}^{N_i} (a_{il} - \alpha)^2 + \lambda_2 \sum
_{i=1}^n \theta_i^2,
\end{equation}
where $\lambda_1 = \sigma_\varepsilon^2/\sigma_a^2$, $\lambda_2 =
\sigma_\varepsilon^2/\sigma_\theta^2$, and $\widetilde
X_{il}(\cdot)$ is the trajectory determined by $a_{il}$, $\theta_i$
and $\bolds\beta$. This can be viewed as a hierarchical maximum
likelihood approach [\citet{Lee2006}], which is considered to be a
convenient alternative to the full (restricted) maximum likelihood
approach. Define
\[
\ell_{ilj}(a_{il},\theta_i,\bolds\beta) :=[Y_{ilj} -
\widetilde X_{il}(t_{ilj};a_{il},\theta_i,\bolds{\beta})]^2 +
\lambda_1
(a_{il} - \alpha)^2/m_{il} + \lambda_2
\theta_i^2\Big/\sum_{l=1}^{N_i}m_{il}.
\]
Then the loss function in (\ref{eq:log-like}) equals $\sum_{i=1}^n
\sum_{l=1}^{N_i}\sum_{j=1}^{m_{il}}\ell_{ilj}(a_{il},\theta
_i,\bolds{\beta})$.
Note that the above distributional assumptions are simply working
assumptions, since the expression in (\ref{eq:log-like}) can also
be viewed as a regularized $\ell_2$ loss with penalties on the
variability of $\bolds\theta$ and $\bs{a}$.

In many problems there are natural constraints on the gradient
function~$g$. Some of these constraints can be expressed in the form
of quadratic constraints in certain derivatives of $g$. Thus, to add
flexibility to our estimation procedure, we allow for incorporating
penalties of the form: $\bolds\beta^T \mathbf{B} \bolds\beta$ for
an $M
\times M$ positive semi-definite matrix $\mathbf{B}$ in the loss
function. Consequently, the modified objective function becomes
%
\begin{equation}\label{eq:objective}
L(\bs{a},\bolds\theta,\bolds\beta) :=\sum_{i=1}^n
\sum_{l=1}^{N_i}
\sum_{j=1}^{m_{il}}\ell_{ilj}(a_{il},\theta_i,\bolds\beta) +
\bolds\beta^T \mathbf{B} \bolds\beta.
\end{equation}
The proposed estimator is then the minimizer of the objective
function:
%
\begin{eqnarray}\label{eq:estimator}
(\widehat{\bs{a}}, \widehat{\bolds{\theta}},
\widehat{\bolds{\beta}}) :=\operatorname{arg}\min_{\bs{a},\bolds\theta
,\bolds\beta}L(\bs{a},\bolds\theta,\bolds\beta).
\end{eqnarray}
Note that here our main interest is the gradient function $g$.
Thus, estimating the parameters of the dynamical system together with
the sample trajectories and their derivatives simultaneously is the
most efficient. In contrast, in a two-stage approach, the
trajectories and their derivatives are first obtained via
pre-smoothing [see, e.g., \citeauthor{ChenWu2008a}
(\citeyear{ChenWu2008a,ChenWu2008b});
\citet{Varah1982}], and then they are used in a nonparametric regression
framework to derive an estimate of $g$. This is inefficient since
estimation errors introduced in the pre-smoothing step effectively
cause a loss of information. Indeed, simulation studies carried out
in Section \ref{sec:simulation} and the supplementary material
[\citet{Paul-supp2011}] show that two-stage estimators suffer from
significant biases in estimating the gradient function $g$.
Alternative ways of estimating $g$ include using the reproducing
kernel Hilbert space framework [\citet{Gu2002}], and controlling the
degree of smoothness of the fitted $g$ by tuning a roughness
penalty.

In the following, we propose a numerical procedure
for solving (\ref{eq:estimator}) that has two main ingredients:
\begin{itemize}
\item
Given $(\bs{a},\bolds\theta,\bolds\beta)$, reconstruct the trajectories
$\{\widetilde X_{il}(\cdot)\dvtx  l=1,\ldots ,N_i\}_{i=1}^n$ and their
derivatives. This can be implemented using a numerical ODE solver,
such as the Runge--Kutta method [\citet{Tenenbaum1985}].

\item
Minimize (\ref{eq:objective}) with respect to
$(\bs{a},\bolds\theta,\bolds\beta)$. This amounts to a nonlinear
regression problem. It can be carried out using either a specialized
nonlinear least squares solver, like the Levenberg--Marquardt method,
[\citet{NocedalWright2006}] or a general optimization procedure,
such as the Newton--Raphson algorithm.
\end{itemize}
The above fitting procedure bears similarity to the local, or
gradient-based, methods discussed by \citet{Li2002},
\citet{Guedj2007} and \citet{Miao2009}, even though their works
focus on parametric ODEs. The main distinction of the proposed
framework and those of \citet{Li2002} and \citet{Guedj2007} lies in
that, for the current setting, the complexity of the model is
allowed to grow with increasing sample size and one eventually needs
to adopt a model selection procedure to select an appropriate model
(as is done in this paper). From purely a model-fitting point of
view, nonlinear mixed-effects (NLME) model-based estimation
procedures may be used in principle to fit each of these parametric
submodels. The work of \citet{Ke2001} on semiparametric
mixed-effects model fitting also shares some common computational
challenges with our model. However, unlike in \citet{Ke2001}, in
our case, the likelihood for the nonparametric component (i.e., the
gradient function) is not available in closed form.

We now briefly describe an optimization procedure based on the idea
of the Levenberg--Marquardt method by linearization of $\{\widetilde
X_{il} (\cdot)\}$ with respect to $a_{il}$, $\theta_i$ and
$\bolds\beta$. We break the updating step into three parts
corresponding to the three different sets of parameters. For each
set of parameters, we first derive a first order Taylor expansion of
the curves $\{\widetilde X_{il}\}$ around their current values and
then update them by a least squares fitting, while keeping the other
two sets of parameters fixed at the current values. This process is
repeated until convergence.

For notational convenience, denote the current estimates by\vspace*{1pt}
$\bs{a}^* :=\{a_{il}^*\}$, $\bolds{\theta}^* :=\{\theta_i^*\}$ and
$\bolds{\beta}^*$, and define the current residuals as
$\tilde\varepsilon_{ilj} := Y_{ilj} - \widetilde
X_{il}(t_{ilj};a_{il}^*,\allowbreak\theta_i^*,\bolds{\beta}^*)$. For each
$i=1,\ldots , n$, and $l=1,\ldots ,N_i$, define the $m_{il} \times1$
column vectors
\[
J_{il,a_{il}^*} :=  \biggl(\frac{\partial}{\partial a_{il}} \widetilde
X_{il}(t_{ilj};a_{il}^*,\theta_i^*,\bolds{\beta}^*) \biggr)_{j=1}^{m_{il}},
 \qquad  \widetilde{\bolds\varepsilon}_{il}= (\widetilde
\varepsilon_{ilj} )_{j=1}^{m_{il}}.
\]
For each $i=1,\ldots ,n$, define the $m_{i\cdot} \times1$ column
vectors
\[
J_{i,\theta_i^*} :=  \biggl(\frac{\partial}{\partial\theta_i}
\widetilde X_{il}(t_{ilj};a_{il}^*,\theta_i^*,\bolds{\beta}^*)
 \biggr)_{j=1, l=1}^{m_{il},N_i};  \qquad  \widetilde{\bolds\varepsilon}_{i}
=  (\widetilde\varepsilon_{ilj} )_{j=1, l=1}^{m_{il},N_i},
\]
where $m_{i\cdot}: =\sum_{l=1}^{N_i} m_{il}$ is the total number of
measurements for the $i$th subject. For each $k=1,\ldots , M$, define
the $m_{\cdot\cdot} \times1$ column vectors
\[
J_{\beta_k^*} :=  \biggl(\frac{\partial}{\partial\beta_k}
\widetilde
X_{il}(t_{ilj};a_{il}^*,\theta_i^*,\bolds{\beta}^*)
\biggr)_{j=1,l=1,i=1}^{m_{il},N_i,n};
 \qquad  \widetilde{\bolds\varepsilon}= (\tilde\varepsilon
_{ilj} )_{j=1,l=1,i=1}^{m_{il},N_i,n},
\]
where $m_{\cdot\cdot} := \sum_{i=1}^n\sum_{l=1}^{N_i} m_{il}$ is the
total number of measurements. Note that, given $\bs{a}^*$,
$\bolds{\theta}^*$ and $\bolds{\beta}^*$, the trajectories $\{
\widetilde
X_{il}(\cdot)\}$'s and their gradients (as well as Hessians) can be
easily evaluated on a fine grid by using numerical ODE solvers such
as the fourth order Runge--Kutta method [see \citet{Paul-supp2011}
for details]. Since, given the trajectories, their gradients satisfy
linear differential equations, the solution may also be obtained
explicitly (see the \hyperref[appm]{Appendix}). The equation for updating $\bolds\beta$,
while keeping $\bs{a}^*$ and $\bolds{\theta}^*$ fixed, is
\[
 [J_{\bolds{\beta}^*}^T J_{\bolds{\beta}^*} + \lambda_3
\operatorname{diag}(J_{\bolds{\beta}^*}^T J_{\bolds{\beta}^*}) +
\mathbf{B} ](\bolds\beta- \bolds\beta^*) = J_{\bolds\beta^*}^T
\widetilde{\bolds\varepsilon} - \mathbf{B}\bolds\beta^*,
\]
where $J_{\bolds\beta^*}:=(J_{\beta_1^*}: \cdots: J_{\beta_M^*})$ is an
$m_{\cdot\cdot} \times M$ matrix. Here $\lambda_3$ is a sequence of
positive constants decreasing to zero as the number of iterations
increases. They are used to avoid possible singularities in the
system of equations. The normal equation for updating $\theta_i$ is
\[
(J_{i,\theta_i^*}^T J_{i,\theta_i^*} +
\lambda_2)(\theta_i-\theta_i^*) = J_{i,\theta_i^*}^T
\widetilde{\bolds\varepsilon}_i - \lambda_2 \theta_i^*,
 \qquad i=1,\ldots ,n.
\]
After updating $\theta_i$'s, we re-center the current estimates such
that their mean is set to be zero. This also helps in stabilizing
the algorithm. The equation for updating $a_{il}$, while keeping
$\theta_i$ and $\bolds{\beta}$ fixed at $\theta_i^*$, $\bolds
{\beta}^*$
is
\[\label{eq:a_diff}
(J_{il,a_{il}^*}^T J_{il,a_{il}^*} +
\lambda_1)(a_{il}-a_{il}^*) = J_{il,a_{il}^*}^T \widetilde{\bolds
\varepsilon}_{il} + \lambda_1 \alpha^*_{il},  \qquad l=1,\ldots , N_i,
\ i=1,\ldots ,n,
\]
where $\alpha^* := \sum_{i=1}^n
\sum_{l=1}^{N_i}a_{il}^*/N_{\cdot}$,   $\alpha_{il}^* = \alpha^* -
a_{il}^*$ with $N_{\cdot}:=\sum_{i=1}^n N_i$ being the total number
of sample curves. Note that on convergence, $\widehat\alpha:=
\alpha^*$ provides an estimate of $\alpha$. The initial estimates
can be conveniently chosen. For example, $a_{il}^{\mathrm{ini}}=Y_{il1}$ and
$\theta_i^{\mathrm{ini}} \equiv0$.

This procedure is quite stable and robust to the initial parameter
estimates. However, it converges slowly in the neighborhood of the
minima of the objective function as it is a first order procedure.
On the contrary, the Newton--Raphson algorithm has a fast convergence
when starting from estimates that are already near the minima. Thus,
in practice, one could first use the above approach (referred to as
the Levenberg--Marquardt step hereafter) to obtain a reasonable
estimate and then use the Newton--Raphson algorithm to expedite the
search of the minima. The derivation of the Newton--Raphson algorithm
is rather standard and thus is omitted. If the true gradient
function $g$ has high complexity, and/or if either the $\theta_i$'s
or the noise are highly variable, the Newton--Raphson algorithm may
be unstable, particularly when the initial conditions
$\bs{a}=\{X_{il}(0)\}$ are also estimated. Under such situations, we
recommend using a (relatively) large number of Levenberg--Marquardt
steps, followed by a one-step Newton--Raphson update.

Note that the tuning parameter $\lambda_3$ plays a different role
than the penalty parameters $\lambda_1$ and $\lambda_2$. The
parameter $\lambda_3$ is used to stabilize the updates of~$\bolds\beta$
and thereby facilitate convergence. Thus, it needs to decrease to
zero with increasing iterations in order to avoid introducing bias
in the estimate. In this paper, we simply set $\lambda_{3j} =
\lambda_3^0 / j$ for the $j$th iteration, for some pre-specified
$\lambda_3^0 > 0$. On the other hand, $\lambda_1$ and $\lambda_2$
are parts of the loss function (\ref{eq:objective}). Their main role
is to control the bias-variance trade-off of the estimators, even
though they also help in regularizing the optimization procedure.
From the likelihood viewpoint, $\lambda_1$ and $\lambda_2$ are
determined by the variances $\sigma_\varepsilon^2$, $\sigma_a^2$ and
$\sigma_\theta^2$ through $\lambda_1 =
\sigma_\varepsilon^2/\sigma_a^2$ and $\lambda_2 =
\sigma_\varepsilon^2/\sigma_\theta^2$. We can estimate these
variances from the current residuals and current values of
$\mathbf{a}$ and $\bolds{\theta}$. By assuming that $m_{il}
> 2$ for each pair $(i,l)$,
\begin{eqnarray*}
\widehat\sigma_\varepsilon^2 &=&
\frac{1}{m_{\cdot\cdot}-N_{\cdot}-n-M}\sum_{i=1}^n
\sum_{l=1}^{N_i}\sum_{j=1}^{m_{il}} (\widetilde
\varepsilon_{ilj})^2,\\
\widehat\sigma_a^2 &=& \frac{1}{N_{\cdot}-1} \sum_{i=1}^n
\sum_{l=1}^{N_i} (a_{il}^* - \alpha^*)^2, \qquad
\widehat\sigma_\theta^2 = \frac{1}{n-1}\sum_{i=1}^n (\theta_i^*)^2.
\end{eqnarray*}
We can then plug in the estimates $\widehat\sigma_\varepsilon^2$,
$\widehat\sigma_a^2$ and $\widehat\sigma_\theta^2$ to get new
values of $\lambda_1$ and $\lambda_2$ for the next iteration.
Instead, if we take the penalized loss function viewpoint, we can
simply treat $\lambda_1$ and $\lambda_2$ as fixed regularization
parameters which can be chosen by model selection criteria (see
Section \ref{sec:model_selection}). Henceforth, we refer to the
method as \texttt{adaptive} if $\lambda_1$ and $\lambda_2$ are
updated after each iteration, and as \texttt{nonadaptive} if they
are kept fixed throughout the optimization.

\subsection{Standard error of the estimates}
\label{subsec:SE}

It is important to obtain the standard error of the estimated
gradient function. Since it is typically not possible to obtain an
estimate of the bias for a nonparametric procedure, we ignore the
bias term and use the best projection of true $g$ in the model space
as the surrogate center (this is the standard practice in
nonparametric literature). Thus, equivalently, we provide an estimate
of the asymptotic variance of $\widehat{\bolds\beta}$. Based on the
asymptotic analysis presented in \citet{Paul2009}, we derive the
following estimate:
%
\begin{equation}\label{eq:var_beta_hat_random}
V(\widehat\beta) :=
\widehat{E}[(\widehat{\bolds\beta}-\bolds\beta)(\widehat{\bolds
\beta}-\bolds\beta)^T]
= \widehat\sigma_\varepsilon^2 \mathbf{W}_n,
\end{equation}
with $\mathbf{W}_n = (\mathbf{A}_n + \mathbf{B} -
\mathbf{C}_n^T(\mathbf{D}_n + \lambda_2
I_n)^{-1}\mathbf{C}_n)^{-1}$, where, $I_n$ is the $n\times n$
identity matrix, $\mathbf{A}_n = {\cal
G}_{\beta\beta}(\widehat{\bolds\theta},\widehat{\bolds\beta})$,
$\mathbf{C}_n = {\cal
G}_{\theta\beta}(\widehat{\bolds\theta},\widehat{\bolds\beta})$,
$\mathbf{D}_n = {\cal
G}_{\theta\theta}(\widehat{\bolds\theta},\widehat{\bolds\beta
})$; where
\[
{\cal G}_{\beta\beta}(\bolds{\theta},\bolds{\beta}) :=
\sum_{i=1}^n\sum_{l=1}^{N_i} \sum_{j=1}^{m_{il}}
 \biggl(\frac{\partial X_{il}}{\partial
\bolds{\beta}}(t_{ilj};a_{il},\theta_i,\bolds{\beta}) \biggr)
 \biggl(\frac{\partial X_{il}}{\partial
\bolds{\beta}}(t_{ilj};a_{il},\theta_i,\bolds{\beta}) \biggr)^T;
\]
${\cal G}_{\theta\beta}(\bolds{\theta},\bolds{\beta})$ is the $n
\times M$
matrix with the $i$th row being
\[
\sum_{l=1}^{N_i} \sum_{j=1}^{m_{il}} \frac{\partial X_{il}}{\partial
\theta_i}(t_{ilj};a_{il},\theta_i,\bolds{\beta})  \biggl(\frac
{\partial
X_{il}}{\partial
\bolds{\beta}}(t_{ilj};a_{il},\theta_i,\bolds{\beta})
\biggr)^T, \qquad i=1,\ldots ,n;
\]
and ${\cal G}_{\theta\theta}(\bolds{\theta},\bolds{\beta})$ is
the $n
\times n$ diagonal matrix with the $i$th diagonal entry
\[
\sum_{l=1}^{N_i} \sum_{j=1}^{m_{il}}  \biggl(\frac{\partial
X_{il}}{\partial
\theta_i}(t_{ilj};a_{il},\theta_i,\bolds{\beta}) \biggr)^2,
 \qquad i=1,\ldots ,n.
\]
Note that the matrices $\mathbf{A}_n$, $\mathbf{C}_n$ and
$\mathbf{D}_n$ are obtained as byproducts of the estimation
procedure. An estimate of the standard error of $\widehat g(x)$ for
$x$ in the domain of $\{\phi_{k,M}\}_{k=1}^M$ is therefore given by
%
\begin{equation}\label{eq:SE_g_hat}
\widehat{\operatorname{SE}}(\widehat g(x)) =  [\bolds\phi_M(x)^T
V(\widehat{\bolds\beta}) \bolds\phi_M(x) ]^{1/2},
\end{equation}
where $\bolds\phi_M(x) := (\phi_{1,M}(x),\ldots ,\phi_{M,M}(x))^T$ and
$V(\widehat{\bolds\beta})$ is as in (\ref{eq:var_beta_hat_random}).
Note that, in the given asymptotic framework, we treat $\theta_i$'s
as random effects and the initial conditions $\{a_{il}\}$ are
assumed to be known. In deriving (\ref{eq:var_beta_hat_random}), we
have ignored the correlation structure between $\theta_i$ and the
gradient of the objective function with respect to $\theta_i$, which
yields a slightly conservative (i.e., upwardly biased) estimate of
the standard error. Obtaining the asymptotic standard error
estimates when the initial conditions $\{a_{il}\}$ are estimated
from the data is beyond the scope of this paper.

As an alternative way of estimating the standard error, one may also
use bootstrap where we resample the sample trajectories
corresponding to each subject, in order to retain the overall
structure of the model. The corresponding bootstrap estimates,
though simple to obtain, are computationally expensive and we do not
pursue this in this paper.

\subsection{Model selection}
\label{sec:model_selection}

After specifying a scheme for the basis func-\break{}tions~$\{\phi_{k,M}(\cdot)\}$,
we still need to determine various model
parameters such as the number of basis functions $M$, the knot
sequence, penalty parameters, etc. In the literature, AIC/BIC/AICc
criteria have been proposed for model selection of parametric
dynamical systems; see, for example, \citet{Miao2009}. Here we
propose an approximate leave-one-curve-out cross-validation score
for model selection. Under the current context, the
leave-one-curve-out CV score can be defined as
%
\begin{equation}\label{eq:CV}
\mathit{CV} := \sum_{i=1}^n\sum_{l=1}^{N_{i}}\sum_{j=1}^{m_{il}}
\ell_{ilj}^{cv}\bigl(\widehat a_{il}^{(-il)}, \widehat
\theta_{i}^{(-il)}, \widehat{\bolds\beta}^{(-il)}\bigr),
\end{equation}
where $\widehat\theta_{i}^{(-il)}$ and
$\widehat{\bolds\beta}^{(-il)}$ are estimates of $\theta_{i}$ and
$\bolds\beta$, respectively, based on the data after dropping the $l$th
curve of the $i$th subject;\vspace*{-1pt} and $\widehat a_{il}^{(-il)}$ is the
minimizer of $\sum_{j=1}^{m_{il}} \ell_{ilj}(a_{il},\widehat
\theta_{i}^{(-il)}, \widehat{\bolds\beta}^{(-il)})$ with respect to
$a_{il}$; and $\ell_{ilj}^{cv}(a_{il},\theta_i,\bolds\beta) := (Y_{ilj}
- \widetilde X_{il}(t_{ilj};a_{il},\theta_i,\bolds\beta))^2$ is the
prediction error loss. When the initial conditions $a_{il} =
X_{il}(0)$ are i.i.d. random variables and are known (and thus we
set $\widehat a_{il}^{(-il)} = a_{il}$), the leave-one-curve-out CV
score gives an asymptotically unbiased estimator of the prediction
error. Calculating CV score (\ref{eq:CV}) is computationally very
demanding because one needs to obtain $\widehat\theta_{i}^{(-il)}$
and~$\widehat{\bolds\beta}^{(-il)}$ for every pair of $(i,l)$.
Therefore, we propose to approximate $\widehat\theta_{i}^{(-il)}$
and $\widehat{\bolds\beta}^{(-il)}$ through a first order Taylor
expansion around the estimates $\widehat\theta_{i},
\widehat{\bolds\beta}$ based on the full data. We then obtain an
approximation of $\widehat a_{il}^{(-il)}$ by minimizing the
corresponding criterion with the approximations of $\widehat
\theta_{i}^{(-il)}$ and $\widehat{\bolds\beta}^{(-il)}$ imputed.
Consequently, we derive\vspace*{1pt} an approximate CV score $\widetilde{\mathit{CV}}$ by
plugging these approximations in (\ref{eq:CV}), which is
computationally inexpensive since all the quantities involved in
computing $\widetilde{\mathit{CV}}$ are byproducts of the Newton--Raphson step
used in model fitting. This approximation scheme is similar to the
one taken in \citet{PengPaul2009} under the context of functional
principal component analysis, which itself is motivated by the work
of \citet{Burman1990}. Detailed derivations are given in the
\hyperref[appm]{Appendix}.

\section{Simulation}\label{sec:simulation}\label{sec4}

\begin{figure}

\includegraphics{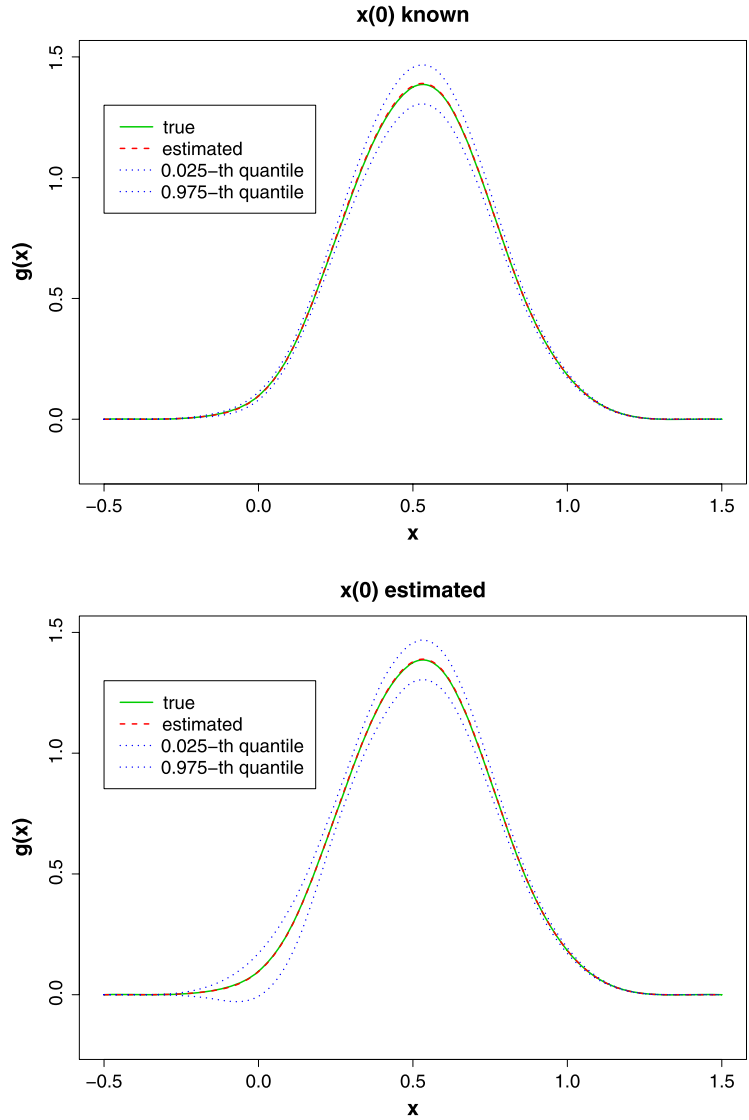}

\caption{True and fitted gradient function $g$ by hierarchical
likelihood approach for the
\texttt{sparse} case. The true model (with $M=4$ B-spline basis
functions with equally spaced knots) is used in fitting. Top panel:
initial conditions $\bs{a}$ are known. Bottom panel: initial
conditions $\bs{a}$ are estimated.} \label{figure:g_fit_band_sparse}
\end{figure}

In this section we conduct a simulation study to demonstrate the
effectiveness of the proposed estimation and model selection
procedures. Since we apply our method to study the plant growth
dynamics in Section \ref{sec:plant}, we consider a simulation
setting that partly mimics that data set. In the simulation, the
true gradient function $g$ is represented by $M_*= 4$ cubic
$B$-spline basis functions with knots at $(0.35,0.6,0.85,1.1)$ and
basis coefficients $\bolds\beta= (0.1, 1.2, 1.6, 0.4)^T$. It is
depicted by the solid curve in Figure
\ref{figure:g_fit_band_sparse}. We consider two different settings
for the number of measurements per curve: \texttt{moderate} case---$m_{il}$'s are independently and identically distributed as
Uniform$[5, 20]$; \texttt{sparse} case---$m_{il}$'s are
independently and identically distributed as Uniform$[3, 8]$.
Measurement times $\{t_{ilj}\}$ are independently and identically
distributed as Uniform$[0,1]$. The scale parameters $\theta_i$'s are
randomly sampled from $N(0,\sigma_\theta^2)$ with
$\sigma_\theta=0.1$; and the initial conditions $a_{il}$'s are
randomly sampled from a $c_a \chi_{k_a}^2$ distribution (to ensure
positivity as well as to study model robustness), with $c_a, k_a >
0$ chosen such that $\alpha=0.25, \sigma_a=0.05$. Finally, the
residuals $\varepsilon_{ilj}$'s are randomly sampled from
$N(0,\sigma_\varepsilon^2)$ with $\sigma_\varepsilon=0.01$.
Throughout the simulation, we set the number of subjects $n=10$ and
the number of curves per subject $N_i \equiv N=20$. Observations
$\{Y_{ilj}\}$ are generated using the model specified by
equations~(\ref{eq:basic}), (\ref{eq:scale}) and (\ref{eq:data_model}). For
all settings, $50$ independent data sets are used to evaluate the
performance of the proposed procedure. The sample trajectories are
evaluated using the 4th order Runge--Kutta method [as described in
\citet{Paul-supp2011}] on an equally spaced grid with grid spacings
$h=0.0005$.

In the estimation procedure, we consider cubic $B$-spline basis
functions with knots at $0.1 + j/M$, $j=1,\ldots ,M$, to model $g$,
where $M$ varies from 2 to 6. Note that here $M=4$ corresponds to
the true gradient function. The Levenberg--Marqardt step is chosen to
be \texttt{nonadaptive}, and the Newton--Raphson step is chosen to
be \texttt{adaptive} (see Section \ref{subsec:modelfit} for the
definition of \texttt{adaptive} and \texttt{nonadaptive}). We
examine three different sets of initial values for $\lambda_1$ and
$\lambda_2$: (i) $\lambda_1=\sigma_{\varepsilon}^2/\sigma_a^2=0.04,
\lambda_2=\sigma_{\varepsilon}^2/\sigma_{\theta}^2=0.01$ (``true''
values); (ii) $\lambda_1=0.01, \lambda_2=0.0025$ (``deflated''
values); (iii) $\lambda_1=0.16, \lambda_2=0.04$ (``inflated''
values). It turns out that the estimation and model selection
procedures are quite robust to the initial choice of $(\lambda_1,
\lambda_2)$, thereby demonstrating the effectiveness of the
\texttt{adaptive} method used in the Newton--Raphson step. Thus, in
the following, we only report the results when the ``true'' values
are used.

%
\begin{table}
\tabcolsep=0pt
\caption{Convergence and model selection based on $50$ independent replicates}
\label{table:selection}
\begin{tabular*}{\textwidth}{@{\extracolsep{\fill
}}lcd{2.0}d{2.0}d{2.0}d{2.0}cd{2.0}cd{2.0}d{2.0}c@{}}
\hline
&& \multicolumn{5}{c}{$\bs{a}$ \textbf{known}} & \multicolumn
{5}{c@{}}{$\bs{a}$ \textbf{estimated}}\\[-5pt]
&& \multicolumn{5}{c}{\hrulefill} & \multicolumn{5}{c@{}}{\hrulefill
}\\
&\textbf{Model}& \multicolumn{1}{c}{$\bolds2$} & \multicolumn
{1}{c}{$\bolds3$} & \multicolumn{1}{c}{\textbf{\textit{4}}} & \multicolumn
{1}{c}{$\bolds5$} & \multicolumn{1}{c}{$\bolds6$} & \multicolumn
{1}{c}{$\bolds2$}
& \multicolumn{1}{c}{$\bolds3$} & \multicolumn{1}{c}{\textbf{\textit{4}}} &
\multicolumn{1}{c}{$\bolds5$} & \multicolumn{1}{c@{}}{$\bolds6$}
\\\hline
\texttt{moderate}& Number converged & 50 & 50 & 50 & 50 & 50 & 50 & 7
& 50 & 50 & 46\\
& Number selected & 0 & 0 & 46 & 1 & \hphantom{5}3 & 0 & 0 & 49 & 1 &
\hphantom{3}0\\[3pt]
\texttt{sparse} & Number converged & 50 & 50 & 50 & 50 & 50 & 50 & 5 &
49 & 44 & 38\\
& Number selected & 0 & 0 & 45 & 0 & \hphantom{5}5 & 1 & 0 & 47 & 1 &
\hphantom{3}1\\
\hline
\end{tabular*}
\end{table}
%
\begin{table}[b]
\tablewidth=285pt
\caption{Estimation accuracy under the true model\protect\tabnoteref
[*]{tab1}}
\label{table:estimation}
\begin{tabular}{@{}lccccc@{}}
\hline
& & \textbf{MISE}$\bolds{(\widehat{g})}$&\textbf{SD(ISE)}&\textbf
{MSPE}$\bolds{(\widehat{\bolds\theta})}$& \textbf{SD(SPE)}\\
\hline
{$\bs{a}$ known} & \texttt{moderate}& 0.069 & 0.072 & 0.085 & 0.095 \\
& \texttt{sparse}& 0.072 & 0.073 & 0.085 & 0.095 \\[3pt]
{$\bs{a}$ estimated} & \texttt{moderate}& 0.088 & 0.079 & 0.086 &
0.095 \\
& \texttt{sparse}& 0.146 & 0.129 & 0.087 & 0.094\\\hline
\end{tabular}
\tabnotetext[*]{tab1}{All the numbers are multiplied by $100$.}
\end{table}

We also compare results when (i) the initial conditions $\bs{a}$ are
known, and hence not estimated; and (ii) when $\bs{a}$ are
estimated. As can be seen from Table \ref{table:selection}, the
estimation procedure converges well and the true model ($M=4$) is
selected most of the times for all the cases. Mean integrated
squared error (MISE) and Mean squared prediction error (MSPE) and
the corresponding standard deviations, SD(ISE) and SD(SPE), based on
50 independent data sets, are used for measuring the estimation
accuracy of $\widehat{g}$ and $\widehat{\bolds\theta}$, respectively.
Since the true model is selected most of the times, we only report
results under the true model in Table \ref{table:estimation}. As can
be seen from this table, when the initial conditions $\bs{a}$ are
known, there is not much difference in the performance between the
\texttt{moderate case} and the \texttt{sparse} case. On the other
hand, when $\bs{a}$ are estimated, the advantages of having more
measurements become more prominent. We also conduct further
simulation studies (results not reported in details here) to check
the effect of increasing the noise level, as well as the dispersion
of the initial conditions $\bs{a}$. When $\bs{a}$ are known, even
with $\sigma_\varepsilon= 0.05$, the convergence is almost
unaffected, and in about 75\% of the cases the true model ($M=4$) is
selected. Increasing $\sigma_a$ to $0.1$ does affect convergence,
especially for larger $M$. But under this setting, even with
$\sigma_\varepsilon=0.05$, the true model converges in 90\% of the
cases and
is selected to be the best in more than 75\% of the cases. When $\bs{a}$
are estimated, the convergence deteriorates more obliviously under
increased noise levels.

In Figure \ref{figure:g_fit_band_sparse} we have a graphical
comparison of the fits when the initial conditions $\bs{a}$ are
known versus when they are estimated in the \texttt{sparse} case. In
the \texttt{moderate} case, there is very little visual difference
under these two settings. We plot the true $g$ (solid curve), the
pointwise mean of $\widehat g$ (broken curve), and 2.5\% and 97.5\%
pointwise quantiles (dotted curves) under the true model. These
plots show that both fits are almost unbiased. Also, when~$\bs{a}$
are estimated, there is greater variability in the estimated $g$ at
smaller values of $x$, mainly due to a scarcity of data in that
region. Indeed, the larger MISE of the estimator of $g$ when initial
conditions are estimated mainly results from the larger MISE on the
domain of $g$ where there is essentially no observed data. Due to
the extrapolation effect, no method without using true  initial
conditions is expected to work well on such a domain, especially
under a nonparametric setting. This point is illustrated in more
detail later in this section (cf. Table
\ref{table:HL_and_two_stage_locpol_reg_sig_theta_0.1}), as well as
in the supplementary material [\citet{Paul-supp2011}, Section~S3].
Overall, as can be seen from these tables and figures, the proposed
estimation and model selection procedures perform effectively.

%
\begin{table}
\caption{Comparison of estimation accuracy of \textup{two-stage
estimators} (either local quadratic smoothing or
parametric regression using true model in the second stage) with
\textup{hierarchical likelihood estimators} (for the \textup{selected
model}, among models with $M=2,\ldots ,6$ B-spline basis functions)
under~the~\texttt{sparse} case}
\label{table:HL_and_two_stage_locpol_reg_sig_theta_0.1}
\tabcolsep=0pt
\begin{tabular*}{\textwidth}{@{\extracolsep{\fill}}lccccc@{}}
\multicolumn{6}{@{}l@{}}{{Two-stage estimator}}\\
\hline
\textbf{Method} & \textbf{Bandwidths} & \textbf{Summary} \\
\textbf{in stage 2} & \textbf{in stage I} & \textbf{statistics} &
$\bolds{x \in[-0.5,0.2]}$ & $\bolds{x \in(0.2,1]}$ & $\bolds{x \in
(1,1.5]}$ \\
\hline
Local quadratic & Optimal& Mean(ISE$(\widehat g)$) & 3.8${}\times10^7$
& 20.177 & 7.3${}\times10^6$ \\
 \quad smoothing & bandwidths & Median(ISE$(\widehat g)$) & \textbf
{4.1}${}\times10^5$ & \textbf{2.398} & \textbf{1.8}${}\times10^3$ \\
& & (SD(ISE$(\widehat g)$)) & ($2.3\times10^8$) & (330.146) &
($5.1\times10^7$)\\
[4pt]
Regression &Optimal & Mean(ISE$(\widehat g)$) & 27.592 & 28.492 & 0.063
\\
 \quad (true model) & bandwidths & Median(ISE$(\widehat g)$) & \textbf{3.812}
& \textbf{2.094} & \textbf{0.004} \\
& & (SD(ISE$(\widehat g)$)) & (423.283) & (565.281) & (1.249)\\
\hline
\end{tabular*}
\begin{tabular*}{\textwidth}{@{\extracolsep{\fill}}lcccc@{}}
\multicolumn{5}{@{}l@{}}{{Hierarchical likelihood estimator}}\\
\hline
& \textbf{Summary} \\
\textbf{Method} & \textbf{statistics}& $\bolds{x \in[-0.5,0.2]}$ &
$\bolds{x \in(0.2,1]}$ & $\bolds{x \in(1,1.5]}$ \\
\hline
{$\bs{a}$ known} & Mean(ISE$(\widehat g)$) & 0.006 & 0.083 & 0.001 \\
& Median(ISE$(\widehat g)$) & \textbf{0.003} & \textbf{0.041} &
\textbf{0.000}\\
& (SD(ISE$(\widehat g)$)) & (0.009) & (0.106) & (0.002)\\
[4pt]
{$\bs{a}$ estimated} & Mean(ISE$(\widehat g)$) & 0.710 & 0.195 &
0.007\\
& Median(ISE$(\widehat g)$) & \textbf{0.025} & \textbf{0.054} &
\textbf{0.000} \\
& (SD(ISE$(\widehat g)$)) & (4.751) & (0.789) & (0.048)\\
\hline
\end{tabular*}
\end{table}

\begin{figure}[t]

\includegraphics{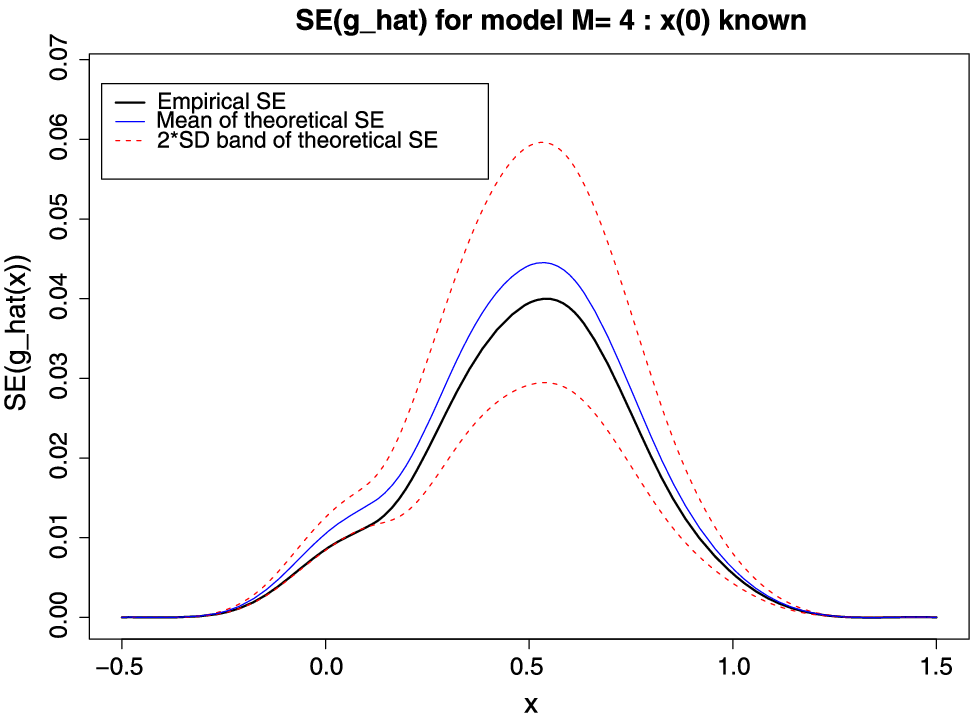}

\caption{Standard error estimates for the simulation study in
Section \protect\ref{sec:simulation}. Pointwise standard error of
$\widehat
g$ for the \texttt{sparse} case with initial conditions $\bs{a}$
known. The true model (with $M=4$ B-spline basis functions) is used
in fitting. Solid black curve: pointwise standard error computed
from 50 replicates. Solid blue curve: averaged pointwise standard
error estimates from (\protect\ref{eq:SE_g_hat}) (based on 50
replicates).
Broken red curve: 2 standard deviations bands for the estimated
pointwise standard error (based on 50 replicates).}\vspace*{-1pt}
\label{figure:g_hat_SE}
\end{figure}

To evaluate the accuracy of the pointwise standard error estimator
given in (\ref{eq:SE_g_hat}), in Figure \ref{figure:g_hat_SE} we
plotted the average of the estimate (blue curve) over 50 independent
data sets and the $\pm2$ standard error bands of the estimates
(broken red curves) based on the same 50 independent data sets under
the true model ($M=4$) when $\bs{a}$ is known. The pointwise
standard errors are also computed empirically from the converged
replicates (black curve) among the 50 simulation runs. We observe
that, although being somewhat conservative, (\ref{eq:SE_g_hat}) gives
a quite satisfactory estimate of the pointwise standard error of~$\widehat
g$.\looseness=1

We also compare the performance of the proposed procedure with a
two-stage approach. Following \citet{ChenWu2008a}, in the first
stage, each individual trajectory $X_{il}(\cdot)$ and its derivative
$X_{il}'(\cdot)$ are estimated by local linear and local quadratic
smoothing, respectively. The bandwidths are chosen by cross-validation.
In the second stage, two different methods for
estimating $g$ are considered with $\{\widehat X_{il}'(t)\}$ as
response and $\{\widehat X_{il}(t)\}$ as predictor: (i) a least
squares regression fit of the basis coefficients using the true
model; (ii) a local quadratic smoothing. A~more detailed
description of the two-stage approach and more simulation studies
are given in the supplementary material [\citet{Paul-supp2011},
Section~S2].

\begin{figure}

\includegraphics{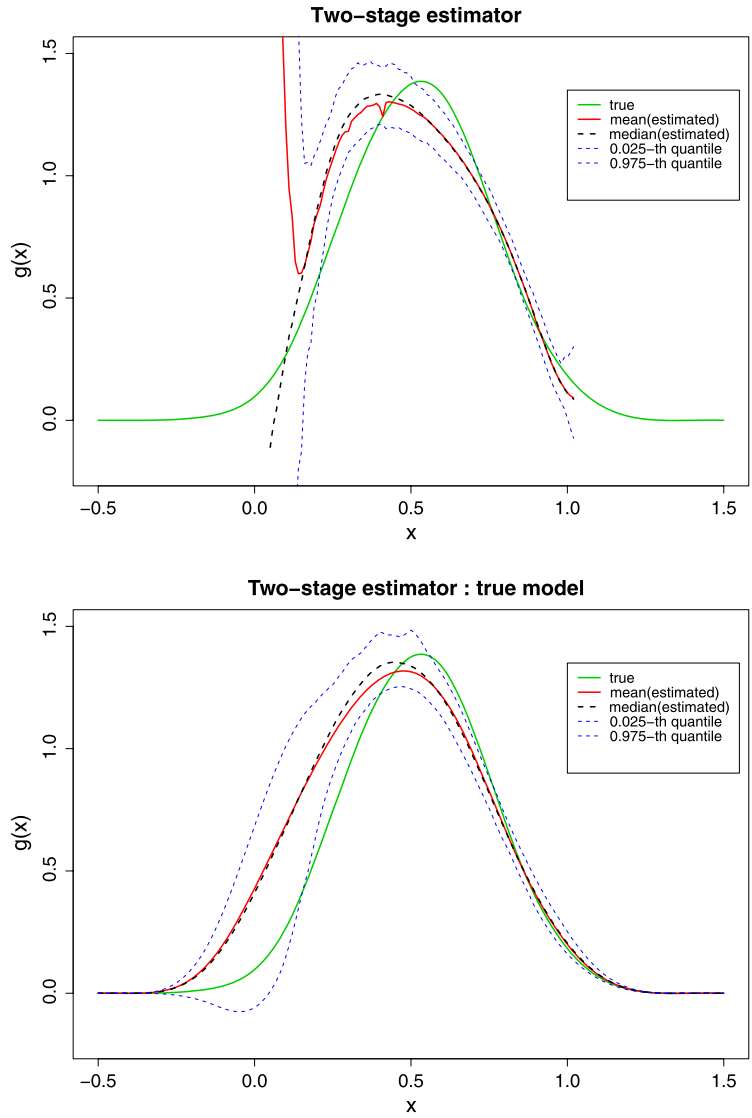}

\caption{True and fitted gradient function $g$ by two-stage approach
for the \texttt{sparse}
case. Top panel: the second stage uses \textup{local quadratic
smoothing}. Bottom
panel: the second stage uses \textup{regression under the true model}.}
\label{figure:bandplot_g_two_stage_sparse_combined_sigtheta0.1}
\end{figure}

In Table \ref{table:HL_and_two_stage_locpol_reg_sig_theta_0.1} we
report the integrated squared errors of the two-stage estimators as
well as those of the hierarchical likelihood estimators (under the
model selected by $\widetilde{\mathit{CV}}$) for the \texttt{sparse} case. While
reporting the risk of the estimators, we divide the domain of $x$
into three regions: $[-0.5,0.2]$, $(0.2,1]$ and $(1,1.5]$. In this
simulation, even though the true gradient function $g$ has support
effectively on $[-0.5,1.5]$, the observed measurements $Y_{ilj}$'s
are almost entirely confined in the region $(0.2,1]$. Due to the
extrapolation effect, methods without using the true initial
conditions are expected to perform (relatively) poorly in the
domains where there is no data. Thus, we divide the domain into
different regions for more informative comparisons across methods.
We also plot the pointwise mean and median and pointwise 95\% bands
around the mean for the two-stage estimators of $g$ in Figure
\ref{figure:bandplot_g_two_stage_sparse_combined_sigtheta0.1}. These
results show that the two two-stage estimators are highly biased and
variable. The one using the true model in the second stage has
better behavior in the regions where there is no data, compared to
the fully nonparametric estimator. However, the level of bias and
variability is much higher than the proposed estimator on all three
regions. Another important observation is that, for the hierarchical
likelihood estimator, the median of integrated squared errors over
the data domain $(0.2,1]$ is comparable for the cases when the
initial condition $\bs{a}$ is known and when $\bs{a}$ is estimated.

To further compare these two approaches, we conduct another
simulation study where all $\theta_i$'s are taken to be zero
(equivalently, $\sigma_\theta= 0$), so that there is no
subject-specific variability. For this simulation, we also consider
a sampling design, referred to as ``very dense,'' in which the
number of measurements per curve is Uniform$[60,100]$ so that the
first stage estimates of the two-stage methods are more accurate.
The number of subjects is chosen to be $n=10$ and there is only one
curve per subject (i.e., $N_i\equiv1$). The results [reported in
Table~S5-5 in \citet{Paul-supp2011}] show that the proposed method
again gives better estimates and it is much less biased (even when
the initial conditions are estimated). The mean integrated squared
error over the data domain $(0.2,1]$ of the hierarchical likelihood
estimator, when $\bs{a}$ is estimated, is much smaller than that of
the two-stage method, even when the true model is used in the second
stage. For a more detailed comparison of the two approaches, see
Section~S2 of \citet{Paul-supp2011}.
Moreover, we also do simulations when the true gradient function $g$
is more complex and does not belong to the model space. The overall
picture for the performance of the proposed estimation and model
selection procedures, as well as the comparison with the two-stage
methods, is consistent with the results presented here. See Section~S3 of \citet{Paul-supp2011} for details.

Finally, we comment on the computational time and the rate of
convergence of the proposed procedure. These depend on several
factors, especially the model complexity and bias, noise level and
criteria for convergence. Typically, the convergence is faster when
$\bs{a}$ is treated as known, as opposed to when it is estimated
from the data. For the simulation study presented here, under the
true model ($M=4$), with $\bs{a}$ known, convergence is generally
achieved in about 30 to 40 Levenberg--Marquardt steps and often in
only 2 to 3 Newton--Raphson steps. The number of Levenberg--Marquardt
steps required for convergence almost doubles when $\bs{a}$ is
estimated. For biased models [including those presented in Section~S3 of \citet{Paul-supp2011}], the convergence often takes more
steps (up to 150 Levenberg--Marquardt steps and several
Newton--Raphson steps). The computational times for the simulation
study presented in this section are summarized in Table
\ref{table:comp_time}. These computations were carried out on a
64-bit Linux machine with Intel Core 2 Quad processors running at
3.2 GHz and with 8~GB~RAM.\looseness=1

\section{Application: Plant growth data}\label{sec:plant}\label{sec5}

In this section we apply the proposed method to the plant growth
data from \citet{Sacks1997} described in the earlier sections. One
goal of this study is to investigate the effect of water stress on
growth displacement rate within the meristem of the primary root of
maize seedlings. Note that, meristem is the tissue in plants
consisting of undifferentiated cells and found in zones of the plant
where growth can take place. The growth displacement rate is defined
as the rate of displacement of a particle placed along the root and
it should not be confused with ``growth rate'' which usually refers
to the derivative of the growth trajectory with respect to time. For
more details, see \citet{Sacks1997}. Growth displacement rate is
important to infer the cell division rate---the local rate of
formation of cells---that is not directly observable in a~changing
population of dividing cells. The growth displacement rate is also
needed for understanding some important physiological processes such
as biosynthesis [\citet{Silk1979}; \citet{Schurr2006}]. Moreover, a useful
growth descriptor called the ``relative elemental growth rate''
(REGR) can be calculated as the gradient of the growth displacement
rate (with respect to distance), which shows quantitatively the
magnitude of growth at each location within the organ.

%
\begin{table}
\tabcolsep=0pt
\caption{Computational cost for the simulation study in
Section \protect\ref{sec:simulation}. Reported quantities are the average
time in seconds and standard deviations (within brackets) over 50
replicates (including the ones~without~convergence)}
\label{table:comp_time}
\begin{tabular*}{\textwidth}{@{\extracolsep{\fill}}lcccccc@{}}
\hline
\multicolumn{2}{c}{\textbf{Model} $\bolds{(M)}$} & \textbf{2} &
\textbf{3} & \textbf{4} & \textbf{5} & \textbf{6} \\
\hline
\texttt{moderate} & $\bs{a}$ known & 11.40 & 20.34 & 28.14 & 41.51 & 42.29 \\
& & (0.24) & (0.68) & (0.73) & (1.53) & (2.39) \\
& $\bs{a}$ estimated & 21.22 & 89.20 & 44.23 & 56.34 & 69.89 \\
& & (1.18) & (18.38) & (4.54) & (9.33) & (23.05) \\
[4pt]
\texttt{sparse} & $\bs{a}$ known & 11.50 & 20.35 & 28.25 & 41.53 & 42.57 \\
& & (0.33) & (0.63) & (0.80) & (1.58) & (3.05) \\
& $\bs{a}$ estimated & 24.01 & 93.58 & 47.06 & 68.57 & 89.57 \\
& & (1.59) & (17.02) & (11.55) & (25.08) & (38.75) \\
\hline
\end{tabular*}
\end{table}

\begin{figure}

\includegraphics{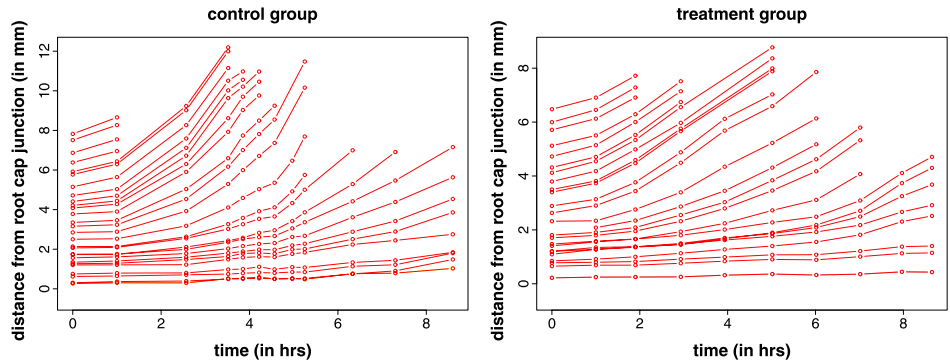}

\caption{Growth trajectories for plant data. Left panel: a plant
in the control group. Right panel: a~plant in the treatment group.}
\label{figure:plant_sample}
\end{figure}

The data consist of measurements on ten plants from a control group
and nine plants from a treatment group where the plants are under
water stress. The meristem region of the root, where the
measurements are taken, is shown in Figure \ref{figure:meristem}
(left panel). The primary roots had grown for approximately $18$
hours in the normal and stressed conditions before the measurements
were taken. The roots were marked at different places using
a~water-soluble marker and high-resolution photographs were used to
measure the displacements of the marked places. The measurements
were in terms of distances from the root cap junction (in
millimeters) and were taken for each of these marked places,
hereafter markers, over an approximate 12-hour period while the
plants were growing. The measurement process is shown schematically
in the right panel of Figure \ref{figure:meristem}. In Figure
\ref{figure:plant_sample} the growth (displacement) trajectories of
one plant with $28$ markers in the control group and another plant
with $26$ markers in the treatment group are depicted. Note that
measurement times are different for these two plants. Also,
measurements were only taken in the meristem. Thus, whenever a marker
grew outside of the meristem, its displacement would not be recorded
at later times anymore. This, together with possible technical
failures (in taking measurements), is the reason that in Figure
\ref{figure:plant_sample} some growth trajectories were cut short.
More sophisticated data acquisition techniques are described in
\citet{Walter2002} and \citet{Basu2007}, where the proposed method
is also potentially applicable.

%
%
%
%

Many studies in plant science such as \citet{Silk1994},
\citet{Sacks1997} and \citet{Fraser1990} all suggest reasonably steady
growth velocity across the meristem under both normal and
water-stress conditions at an early developmental stage. Moreover,
exploratory regression analysis based on empirical derivatives and
empirical fits of the growth trajectories indicates that time is
not a significant predictor and, thus, an autonomous model is
reasonable. This also means that time zero does not play a role in
terms of estimating the dynamical system and there is also no
additional variation associated with individual markers. In
addition, the form of the gradient function $g$ is not known to the
plant scientists, only its behavior at root cap junction and at some
later stage of growth are known [\citet{Silk1994}]. Figure
\ref{figure:emperical_scatter}, the scatter plot of empirical
derivatives versus empirical fits in the treatment group, indicates
that there is an increase in the growth displacement rate starting
from a zero rate at the root cap junction, followed by a nearly
constant rate beyond a certain location. This means that growth
stops beyond this point and the observed displacements are due to
growth in the part of the meristem closer to the root cap junction.
Where and how growth stops is of considerable scientific interest.
These boundary behaviors also imply that a linear ODE model is
obviously not appropriate. In addition, popular parametric models
such as the Michaelis--Menten type either do not satisfy the boundary
constraints and/or have parameters without clear interpretations in
the current context. Moreover, there is some controversy among plant
scientists about the possible existence of a~``growth bump'' in the
middle of the meristem. Taking all these features into
consideration, the semiparametric model proposed in this paper is
appropriate for investigating the scientific questions associated
with this study, in particular, comparing the baseline growth
displacement rates between the treatment and control groups. Notice
that, in order for the proposed estimation method to give an accurate
estimate of the gradient function, we need only that the
measurement on the state variable $x$ is dense in its domain, and
that the measurement errors are independent across time. These are
satisfied for the plant data since, even though each trajectory is
recorded at a relatively small number of time points, there is a
fairly large number of trajectories for each plant, corresponding to
the different initial conditions. Note that, for each plant, the
number of measurements is indeed the sum total of all the
measurements for its  different trajectories. Moreover, the proposed
method combines information across different plants (subjects),
which allows one to fit the model reasonably well even with
relatively few measurements per subject.

Now consider the model described in Section \ref{sec:model}. For the
control group, we have the number of curves per subject $N_i$
varying in between $10$ and~$29$; and for the water stress group, we
have $12 \leq N_i \leq31$. The observed growth displacement
measurements $\{Y_{ilj}\dvtx j=1,\ldots ,m_{il}, l=1,\ldots ,N_i\}_{i=1}^n$
are assumed to follow model (\ref{eq:data_model}), where $m_{il}$ is
the number of measurements taken for the $i$th plant at its $l$th
marker, which varies between $2$ and $17$; and $\{t_{ilj}\dvtx
j=1,\ldots ,m_{il}\}$ are the times of measurements, which are in
between $[0,12]$ hours. Altogether, for the control group there are
$228$ curves with a total of $1\mbox{,}486$ measurements and for the
treatment group there are $217$ curves with $1\mbox{,}712$ measurements in
total. Note that the constraint at the root cap junction
corresponds to $g(0) = 0=g'(0)$, which is imposed by simply omitting
the constant and linear terms in the spline basis. The flatness of
$g$ at a (unknown) distance away from the root cap junction means
that $g'(x) = 0$ for $x \geq A$ for some constant $A > 0$. In order
to impose this, as part of the objective function
(\ref{eq:objective}), we use
\[
\bolds\beta^T \mathbf{B} \bolds\beta:=\lambda_R \int_A^{2A}
(g'(x))^2\,dx
= \lambda_R \bolds\beta^T\biggl [\int_A^{2A} \bolds\phi'(x) (\bolds
\phi'(x))^T
dx\biggr]\bolds\beta,
\]
where $\bolds\phi= (\phi_{1,M},\ldots ,\phi_{M,M})^T$ and $\lambda_R$
is a large positive number quantifying the severity of this
constraint; and $A>0$ determines where the growth displacement
rate becomes a constant. $A$ and $\lambda_R$ are both adaptively
determined by the model selection scheme discussed in Section
\ref{sec:model_selection}. Moreover, since the
initial conditions (marker positions) $\{a_{il}\}$ are chosen
according to some fixed experimental design (though measured with
errors), it is not appropriate to shrink their estimates toward a
fixed number. Hence, we set $\lambda_1 = 0$ in the loss function
(\ref{eq:objective}).

\begin{figure}[b]

\includegraphics{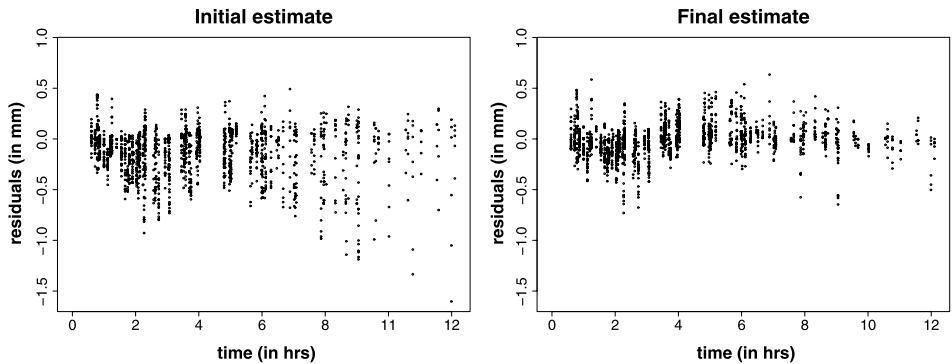}

\vspace*{-2pt}
\caption{Residual versus time plots for the treatment group. Left
panel: fit by \texttt{stepwise-regression}. Right panel: fit by the
proposed method based on maximizing the hierarchical likelihood.}
\label{figure:res_fit_waterstress}
\vspace*{-2pt}
\end{figure}

Before fitting the proposed model, we first describe a simple
regression-based method for getting a crude initial estimate of the
function $g(\cdot)$, as well as selecting a candidate set of knots.
This involves (i) computing the re-scaled empirical derivatives
$e^{-\widehat\theta_i^{(0)}}\widehat{X}^{\prime}_{ilj}$ of the
sample curves from the data, where the empirical derivatives are
defined by taking divided differences:\vspace*{1pt} $\widehat{X}^{\prime}_{ilj}
:= (Y_{il(j+1)}-Y_{ilj})/(t_{il(j+1)}-t_{ilj})$, and
$\widehat\theta_i^{(0)}$ is a preliminary estimate of~$\theta_i$;
and (ii) regressing the re-scaled empirical derivatives onto a set
of basis functions evaluated at the corresponding sample averages:
$\widehat{X}_{ilj} := (Y_{il(j+1)}+Y_{ilj})/2$. In this paper we
use the basis $\{x^2,x^3,(x - x_k)_+^3\}_{k=1}^K$ with a
pre-specified, dense set of knots $\{x_k\}_{k=1}^K$. Then, a model
selection procedure, like the stepwise regression, with either AIC
or BIC criterion, can be used to select a set of candidate knots. In
the following, we shall refer to this method as
\texttt{stepwise-regression}. A similar method is employed by
\citet{Sacks1997}. The resulting estimate of $g$ and the selected
knots can then act as a starting point for the proposed procedure.
We expect this simple method to work reasonably well only when the
number of measurements per curve is moderately large. Comparisons
given later (Figure \ref{figure:res_fit_waterstress}) demonstrate a
clear superiority of the proposed method over this simple approach.

%
\begin{table}[b]
\tabcolsep=0pt
\caption{Model selection for real data. Control group: approximate CV
scores for four \textup{submodels} of the model
selected by the AIC criterion in the \texttt{stepwise-regression}
step. \texttt{M1}:~knots${} = (3.0,
4.0, 5.0, 6.0, 9.0, 9.5)$; \texttt{M2}: knots${} =(3.0, 4.0, 5.5, 6.0,
9.0, 9.5)$;
\texttt{M3}:~knots${} =(3.0, 4.0, 6.0, 9.0, 9.5)$; \texttt{M4}:
knots${} =(3.0, 4.5, 6.0, 9.0, 9.5)$. Treatment group: approximate CV
scores for the model \texttt{M}: knots${} =(3.0, 3.5,
7.5)$}
\label{table:CV_real}
\begin{tabular*}{\textwidth}{@{\extracolsep{\fill}}lcccccc@{}}
\multicolumn{7}{@{}l@{}}{{Control}}\\
\hline
& \multicolumn{3}{c}{$\bolds{\lambda_R = 10^3}$} & \multicolumn
{3}{c@{}}{$\bolds{\lambda_R = 10^5}$}\\[-5pt]
& \multicolumn{3}{c}{\hrulefill} &
\multicolumn{3}{c@{}}{\hrulefill}\\
\textbf{Model} & $\bolds{A = 8.5}$ & $\bolds{A = 9}$ & $\bolds{A =
9.5}$ & $\bolds{A = 8.5}$ & $\bolds{A = 9}$ & $\bolds{A = 9.5}$ \\
\hline
\texttt{M1} & 53.0924 & 53.0877 & 53.1299 & 54.6422 & 53.0803 &
53.1307 \\
\texttt{M2} & 53.0942 & 53.0898 & 53.1374 & 54.5190 & 53.0835 &
53.1375 \\
\texttt{M3} & 53.0300 & 53.0355 & 53.0729 & 53.8769 & \textbf
{53.0063} & 53.0729 \\
\texttt{M4} & 53.0420 & 53.0409 & 53.0723 & 54.0538 & 53.0198 &
53.0722 \\
\hline
\end{tabular*}
\begin{tabular*}{\textwidth}{@{\extracolsep{\fill}}lcccccc@{}}
\multicolumn{7}{@{}l@{}}{{Treatment}}\\
\hline
& \multicolumn{3}{c}{$\bolds{\lambda_R = 10^3}$} & \multicolumn
{3}{c@{}}{$\bolds{\lambda_R = 10^5}$}\\[-5pt]
& \multicolumn{3}{c}{\hrulefill} &
\multicolumn{3}{c@{}}{\hrulefill}\\
\textbf{Model} & $\bolds{A = 7}$ & $\bolds{A = 7.5}$ & $\bolds{A =
8}$ & $\bolds{A = 7}$ & $\bolds{A = 7.5}$
& $\bolds{A = 8}$ \\
\hline
\texttt{M} & \textbf{64.9707} & 64.9835 & 64.9843 &
$65.5798$\tabnoteref[*]{tab2} & 64.9817 & 64.9817 \\
\hline
\end{tabular*}
\tabnotetext[*]{tab2}{No convergence.}
\end{table}

We fit the proposed model to the control group and the treatment
group separately. For the control group, we first use the procedure
described in Section \ref{subsec:modelfit} with $g$ represented in
cubic $B$-splines with $M$ (varying from $2$ to $12$) equally spaced
knots. At this stage, we set $\bolds\beta^{\mathrm{ini}} = 1_M$,
$\bolds\theta^{\mathrm{ini}} = 0_n$, $\bs{a}^{\mathrm{ini}} =
(X_{il}(t_{il1})\dvtx l=1,\ldots ,N_i)_{i=1}^n$. The criterion based on
the approximate CV score [equation (\ref{eq:approx_CV}) in the
\hyperref[appm]{Appendix}] selects the model with $M=9$ basis functions. This is not
surprising since, when equally spaced knots are used, usually a large
number of basis functions are needed to fit the data adequately. In
order to get a more parsimonious model, we consider the
\texttt{stepwise-regression} method to obtain a candidate set of
knots. We use $28$ equally spaced candidate knots on the interval
$[0.5,14]$ and use the fitted values
$\{\widehat{\theta}_i^{(0)}\}_{i=1}^{10}$ from the previous
$B$-spline fit. The AIC criterion selects a model with 10 knots among
these 28 candidate knots, plus the quadratic term. We then
consider various submodels with knots chosen from this set of selected
knots and fit them again using the proposed estimation procedure.
The approximate CV scores for a number of different submodels are
reported in
Table \ref{table:CV_real}. The parameters $A$ and $\lambda_R$ are
also varied and selected by the approximate CV score. Based on the
approximate CV score, the model with knot sequence $(3.0, 4.0, 6.0,
9.0, 9.5)$ and $(A,\lambda_R) = (9,10^5)$ is selected. Also note that
the model
selected by \texttt{stepwise-regression} has a larger CV score than
those of the
models reported in Table \ref{table:CV_real}. A similar procedure is
applied to the treatment group. It turns out that the model with
knot sequence $(3.0, 3.5, 7.5)$, which is also selected by \texttt
{stepwise-regression},
has considerably smaller CV score compared to all other candidate
models, and, hence,
we only report the CV scores under this model in Table \ref
{table:CV_real} with various
choices of $(A, \lambda_R)$. It shows that $(A, \lambda_R) =
(7,10^3)$ has the smallest approximate CV score.

%
\begin{figure}

\includegraphics{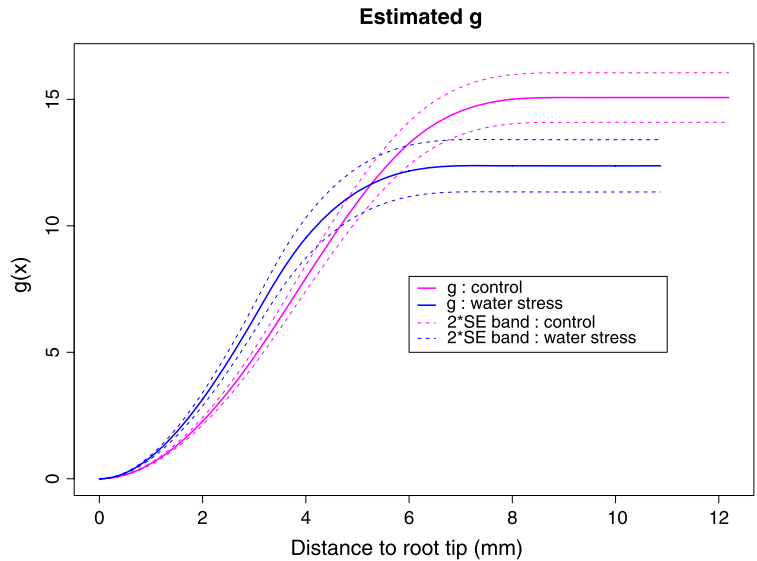}

\caption{Fitted gradient function $\widehat g$, and pointwise 2
standard error bands under the selected models for control and
treatment groups.} \label{figure:compare_link_natural}
\end{figure}

\begin{figure}[t]

\includegraphics{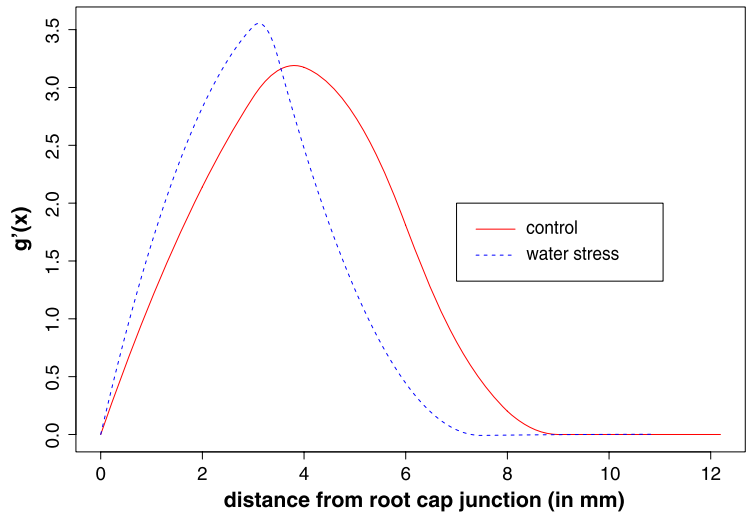}

\caption{Fitted relative elemental growth rate (REGR) under the
selected models for control and treatment groups, respectively. The
REGR is computed by differentiating the estimated gradient function
$g$.} \label{figure:compare_regr_natural}\vspace*{-3pt}
\end{figure}

Figure \ref{figure:compare_link_natural} shows the estimated
gradient functions $\widehat{g}$ under the selected models for the
control and treatment groups, respectively. Apart from $\widehat g$,
we also plot the estimated pointwise two-standard error bands using
(\ref{eq:SE_g_hat}). The fact that the bands are generally
nonoverlapping except for a small region clearly indicates that the
baseline growth displacement rates for the control and treatment
groups are different. The plot also shows that there is no growth
bump for either group. In the part of the meristem closer to the
root cap junction (distance within $\sim$5.5 mm), the growth
displacement rate for the treatment group is higher than that for
the control group. This is probably due to the greater cell
elongation rate under water stress condition in this part of the
meristem so that the root can reach deeper in the soil to get enough
water. This is a known phenomenon in plant science. The growth
displacement rate for the treatment group flattens out beyond
a~distance of about 6 mm from the root cap junction. The same
phenomenon happens for the control group, however, at a further
distance of about 8 mm from the root cap junction. Also, the final
constant growth displacement rate of the control group is higher
than that of the treatment group. This is due to the stunting effect
of water stress on these plants, which results in an earlier stop of
growth and a slower cell division rate. Figure
\ref{figure:compare_regr_natural} shows the estimated relative
elemental growth rates (i.e., $\widehat{g}'$) for these two groups.
Relative elemental growth rate (REGR) relates the magnitude of
growth directly to the location along the meristem. For both groups,
the growth is fastest in the middle part of the meristem ($\sim$3.8
mm for control group and $\sim$3.1 for treatment group), and then
growth dies down pretty sharply and eventually stops. We observe a
faster growth in the part of the meristem closer to the root cap
junction for the water stress group and the growth dies down more
quickly compared to the control group. The shape of the estimated
$g$ may suggest that it might be modeled by a logistic function with
suitably chosen location and scale parameters, even though the
scientific meaning of these parameters is unclear and the boundary
constraints are not satisfied exactly. As discussed earlier, there
is insufficient knowledge from plant science to suggest a functional
form beforehand. This signifies the major purpose and advantage of
nonparametric modeling, which is to provide insights and to suggest
candidate parametric models for further studies.

%
%
\begin{figure}

\includegraphics{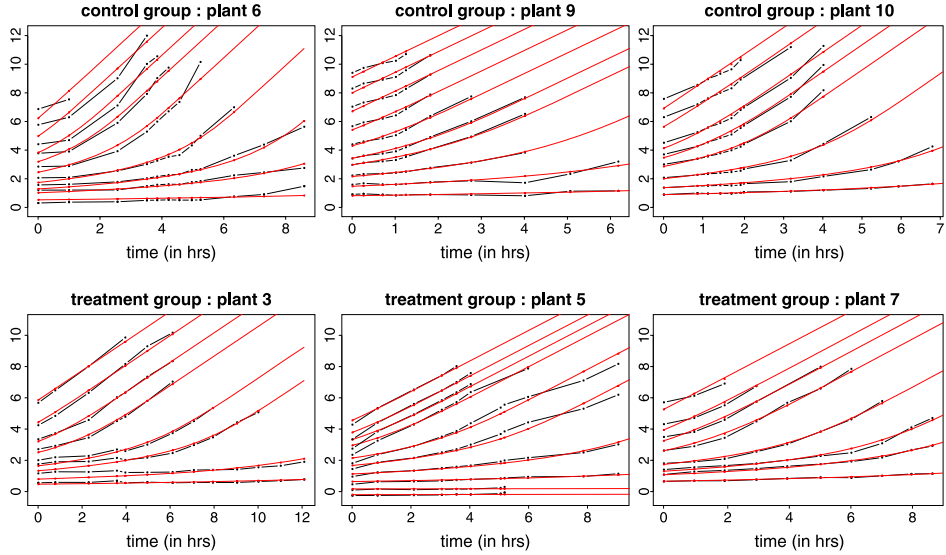}
\vspace*{-2pt}
\caption{Observed (black) and fitted (red) trajectories (under the
selected models) for the plant data. Every third trajectory of each
plant is plotted. Top panel: (from left) plant \# 6, 9, 10 in the
control group. Bottom panel: (from left) plant \# 3, 5, 7 in the
treatment (water stress)
group.}\label{figure:fitted_vs_obs_combined}
\vspace*{-2pt}
\end{figure}
%

In order to check how our method performs in terms of estimating
individual sample trajectories, we solved the differential equation
model for each plant $i$ with fitted values of $X_{il}(0)$,
$\theta_i$ and $g$. Figure \ref{figure:fitted_vs_obs_combined} shows
the fitted (under the selected model) and observed trajectories for
three plants each from the control and the treatment groups. As can
be seen from this figure, although there are subject-specific
variabilities in the fits, the overall shapes of the trajectories
are captured fairly well. Figure \ref{figure:res_fit_waterstress}
shows the residual versus time plot for the treatment group. The
plot for the control group is similar and thus is omitted. This
plot shows that the proposed procedure based on minimizing the objective
function (\ref{eq:objective}) has much smaller and more evenly
spread residuals (SSE $=64.50$) than the fit by
\texttt{stepwise-regression} (SSE $= 147.57$), indicating a clear
benefit of the more sophisticated approach. Overall, the estimation
and model selection procedures give reasonable fits under both
experimental conditions. Note that, for the first six hours, the
residuals (right panel of Figure \ref{figure:res_fit_waterstress})
show some time-dependent pattern, which is not present for later
times. Since throughout the whole $12$ hour period the residuals
remain small compared to the scale of the measurements, the
autonomous system approximation seems to be adequate at least for
practical purposes. Nevertheless, modeling growth dynamics through
nonautonomous systems may enable scientists to determine the stages
of growth that are not steady across a region of the root. This
aspect is discussed briefly in Section \ref{sec:dis}.

\section{Discussion}
\label{sec:dis}

The model and the fitting procedures presented in this paper are
quite flexible and effective in terms of modeling autonomous
dynamical systems nonparametrically when the data are from a number
of subjects and when the underlying population level dynamics is of
interest. When applying the proposed method to the plant growth
data, we obtain results that are scientifically sensible. For the
plant data, $g$ is nonnegative and, thus, a modeling scheme imposing
this constraint may be more advantageous. However, the markers are
all placed at a certain distance from the root cap junction, where
the growth displacement rate is already positive, and the total
number of measurements per plant is moderately large. These mean
that explicitly imposing nonnegativity is not crucial for the plant
data, a fact also supported by the estimates which turn out to be
nonnegative and the simulation results where the resulting
estimators of $g$ are always nonnegative for the \texttt{moderate}
and/or ``$\bs{a}$ known'' cases. In general, if $g$ is strictly
positive (strictly negative) over the domain of interest, then we
can model the logarithm of $g$ (resp., $-g$) by basis
representation.

The proposed approach is flexible in terms of incorporating various
constraints on the dynamics and is able to capture features of the
dynamical system which are not known to us  a priori. It can
also be extended to incorporate covariate effects, as well as to
model nonautonomous systems which are currently under
investigation. Even though in this paper we use the plant growth
data as an illustration, the proposed framework is potentially
useful to many other studies with similar types of data, where
estimating the underlying dynamical system is of interest. For
example, the data set collected as part of the Multicenter AIDS
Cohort Study [\citet{Kaslow1987}; \citet{Diggle2002}] can be used to study
the dynamics of the CD4$+$ counts. Investigating the dynamics of CD4$+$
counts at a population level, while also taking into account
individual effects, is of great importance to understand the
progression of AIDS. This data set consists of 2,376 measurements of
CD4$+$ cell counts against time since seroconversion (time when HIV
becomes detectable) for 369 infected men enrolled in the study. In
this data set, each patient is a subject and there is one sample
curve associated with each subject which reflects CD4$+$ counts over
time. Moreover, each curve is only observed at a few time points and
the set of measurement times is different across patients. The
estimation procedure proposed in this paper can be adjusted
appropriately to deal with such scenarios more effectively.
Specifically, in order to deal with a large number of random
effects, instead of the hierarchical likelihood approach, we can
adopt a marginal maximum likelihood approach. These are topics of
our ongoing research.

\begin{appendix}\label{appm}
\section*{Appendix}

\subsection*{Gradient of the sample trajectories}

Note that $X_{il}(\cdot)$ satisfies
%
\begin{equation}\label{eq:tilde_X_i}
X_{il}(t) = a_{il} + \int_{0}^t e^{\theta_i} \sum_{k=1}^M
\beta_k \phi_{k,M}( X_{il}(s))\,ds,  \qquad t\in[0,1].
\end{equation}
Differentiating (\ref{eq:tilde_X_i}) with respect to the parameters,
we have
\begin{eqnarray*}
X_{il}^{a_{il}}(t) &:=& \frac{\partial X_{il}(t)}{\partial a_{il}}
= 1 + \int_{0}^t \frac{\partial X_{il}(s)}{\partial a_{il}}
e^{\theta_i}
\sum_{k=1}^M \beta_k \phi_{k,M}'( X_{il}(s))\,ds, \\
X_{il}^{\theta_i}(t) &:=& \frac{\partial X_{il}(t)}{\partial
\theta_i}\\
 &=& \int_{0}^t  \Biggl[\frac{\partial X_{il}(s)}{\partial
\theta_i} e^{\theta_i} \sum_{k=1}^M \beta_k \phi_{k,M}'( X_{il}(s)) + e^{\theta_i} \sum_{k=1}^M \beta_k
\phi_{k,M}(X_{il}(s)) \Biggr]\,ds, \\
X_{il}^{\beta_r}(t) &:=& \frac{\partial X_{il}(t)}{\partial\beta_r}
=\!\int_{0}^t \! \Biggl[\frac{\partial X_{il}(s)}{\partial\beta_r}
e^{\theta_i} \sum_{k=1}^M \beta_k \phi_{k,M}'( X_{il}(s)) +
e^{\theta_i} \phi_{r,M}( X_{il}(s)) \Biggr]\,ds
\end{eqnarray*}
for $i=1,\ldots ,n$; $l=1,\ldots , N_i$; $r=1,\ldots ,M$.
In other words, these functions satisfy the
linear differential equations:
\begin{eqnarray}
\frac{d}{dt} X_{il}^{a_{il}}(t) &=& X_{il}^{a_{il}}(t)
e^{\theta_i}\sum_{k=1}^M \beta_k \phi_{k,M}'( X_i(t)),  \qquad
X_{il}^{a_{il}}(0)=1,
\nonumber\\
\frac{d}{dt} X_{il}^{\theta_i}(t) &=& X_{il}^{\theta_i}(t)
e^{\theta_i} \sum_{k=1}^M \beta_k \phi_{k,M}'(
X_{il}(t)) + e^{\theta_i} \sum_{k=1}^M \beta_k \phi_{k,M}(
X_{il}(t)),\nonumber\\[-8pt]
  \eqntext{X_{il}^{\theta_i}(0) = 0,}
\\
\frac{d}{dt} X_{il}^{\beta_r}(t) &=& X_{il}^{\beta_r}(t) e^{\theta_i}
\sum_{k=1}^M \beta_k \phi_{k,M}'( X_{il}(t)) + e^{\theta_i}
\phi_{r,M} (X_{il}(t)),  \qquad  X_{il}^{\beta_r}(0)= 0.\nonumber
\end{eqnarray}
If the $a_{il}$'s are positive and the function $g_{\bolds{\beta}} :=
\sum_{k=1}^M \beta_k \phi_{k,M}$ is positive on the domain of
$a_{il}$'s, then the trajectories $X_{il}(t)$ are nondecreasing in
$t$. In this case, and more generally, whenever the solutions exist
on the time interval $[0,1]$ and $g_{\bolds{\beta}}$ is continuously
differentiable the gradients of the trajectories can be solved
explicitly:
%
\begin{eqnarray}\label{eq:tilde_X_a_i_closed}
X_{il}^{a_{il}}(t) &=& \frac{g_{\bolds{\beta}}
(X_{il}(t))}{g_{\bolds{\beta}}( X_{il}(0))},\\ \label{eq:tilde_X_theta_i_closed}
X_{il}^{\theta_i}(t) &=& e^{\theta_i} t
g_{\bolds{\beta}}( X_{il}(t)),\\\label{eq:tilde_X_beta_closed}
X_{il}^{\beta_r}(t) &=& g_{\bolds{\beta}}( X_{il}(t))
\int_{ X_{il}(0)}^{ X_{il}(t)} \frac{\phi_{r,M}(x)}{(g_{\bolds
{\beta}}(x))^2}\,dx.
\end{eqnarray}
We verify (\ref{eq:tilde_X_beta_closed}). Proofs
(\ref{eq:tilde_X_a_i_closed}) and (\ref{eq:tilde_X_theta_i_closed})
are similar. We can express
\begin{eqnarray}
X_{il}^{\beta_r}(t) &=& e^{\theta_i} \int_{0}^{t}
\phi_{r,M}( X_{il}(s)) \exp \biggl(e^{\theta_i} \int_s^t
g_{\bolds{\beta}}'( X_{il}(u))\,du \biggr)\,ds \nonumber\\
&=& e^{\theta_i} \int_{0}^{t} \phi_{r,M}( X_{il}(s))
\exp \biggl(\int_s^t \frac{g_{\bolds{\beta}}'(
X_{il}(u))}{g_{\bolds{\beta}}( X_{il}(u))} X_{il}'(u)\,du \biggr)\,ds
\nonumber\\
&&  \eqntext{[\mbox{using } X_{il}'(u) = e^{\theta_i}
g_{\bolds{\beta}}( X_{il}(u))]} \\
&=& e^{\theta_i} \int_{0}^{t} \phi_{r,M}( X_{il}(s))
\exp\bigl(\log g_{\bolds{\beta}}( X_{il}(t)) - \log g_{\bolds{\beta}}(
X_{il}(s))\bigr)\,ds\nonumber\\
&=& g_{\bolds{\beta}}( X_{il}(t)) \int_{0}^t \frac{\phi_{r,M}( X_{il}(s))}
{(g_{\bolds{\beta}}(X_{il}(s)))^2} X_{il}'(s)\,ds\nonumber\\
&=& g_{\bolds{\beta}}( X_{il}(t)) \int_{ X_{il}(0)}^{ X_{il}(t)}
\frac{\phi_{r,M}(x)}{(g_{\bolds{\beta}}(x))^2}\,dx.\nonumber
\end{eqnarray}

\subsection*{Derivation of $\widetilde{\mathit{CV}}$}

Observe that, when evaluated at the estimate $\widehat{\bs{a}}$,
$\widehat{\bolds\theta}$ and $\widehat{\bolds\beta}$ based on the full
data,
%
\begin{eqnarray}\label{eq:CV_first}
   \frac{\partial}{\partial\theta_i} \biggl(\sum_{l,j} \ell
_{ilj}^{cv} \biggr)
+2\lambda_2\theta_i&=&0,  \qquad i=1,\ldots ,n;  \nonumber
\\[-8pt]
\\[-8pt]
\frac{\partial}{\partial\bolds\beta} \biggl(\sum_{i,l,j}
\ell_{ilj}^{cv} \biggr)+ 2 \mathbf{B} \bolds\beta&=&0.
\nonumber
\end{eqnarray}
Whereas, when evaluated at the drop $(i,l)$-estimates, $\widehat
a_{il}^{(-il)}, \widehat\theta_{i}^{(-il)},
\widehat{\bolds\beta}^{(-il)}$,
%
\begin{eqnarray}\label{eq:CV_second}
\frac{\partial}{\partial\theta_i} \biggl(\sum_{l^*, j\dvtx  l^* \neq l}
\ell_{i l^* j}^{cv} \biggr)
+ 2 \lambda_2\theta_i&=&0;  \nonumber
\\[-8pt]
\\[-8pt] \frac{\partial}{\partial\bolds\beta
} \biggl(\sum_{i^*,
l^*, j\dvtx  (i^*,l^*) \neq(i,l)} \ell_{i^* l^* j}^{cv}  \biggr)+2 \mathbf
{B} \bolds\beta&=&0.
\nonumber
\end{eqnarray}
Expanding the left-hand side of (\ref{eq:CV_second}) around
$\widehat{\bolds\beta}$ and $\widehat{\bolds\theta}$, and using
(\ref{eq:CV_first}), we obtain the following first order
approximations:
\begin{eqnarray*}\label{eq:CV_est_approx}
\widehat\theta_{i}^{(-il)} &\approx& \widetilde\theta_{i}^{(-il)}
:= \widehat\theta_{i} +
 \Biggl[\sum_{l'=1}^{N_{i}} \sum_{j'=1}^{m_{il'}}\frac{\partial^2
\ell_{il'j'}^{cv}}{\partial\theta_{i}^2}+2\lambda_2  \Biggr]^{-1}
\sum_{j=1}^{m_{il}}  \biggl(\frac{\partial
\ell_{ilj}^{cv}}{\partial\theta_i} \biggr), \\
\widehat{\bolds\beta}^{(-il)} &\approx& \widetilde{\bolds\beta}^{(-il)}
:= \widehat{\bolds\beta}+
\Biggl [\sum_{i'=1}^n \sum_{l'=1}^{N_{i'}} \sum_{j'=1}^{m_{i'l'}}
\frac{\partial^2 \ell_{i'l'j'}^{cv}}{\partial\bolds\beta\,\partial
\bolds\beta^T}+ 2 \mathbf{B} \Biggr]^{-1}  \Biggl(\sum_{j=1}^{m_{il}}
\frac{\partial\ell_{ilj}^{cv}}{\partial\bolds\beta} \Biggr).
\end{eqnarray*}
In the above, the gradients and Hessians of $\ell_{ilj}^{cv}$ are
all evaluated at\vspace*{1pt} $(\widehat{\bs{a}}, \widehat{\bolds\theta},
\widehat{\bolds\beta})$, and, thus, they have already been computed
on a fine grid in the course~of~ob\-taining these estimates. Hence, there
is almost no additional computatio\-nal cost to obtain these
approximations. Now for $i=1,\ldots ,n; l=1,\ldots ,N_i$, define
\[
\widetilde a_{il}^{(-il)}= \arg\min_{a} \sum_{j=1}^{m_{il}}  \bigl[Y_{ilj}
- \widetilde X_{il}\bigl(t_{ilj}; a,\widetilde{\theta_i}^{(-il)},
\widetilde{\bolds{\beta}}^{(-il)}\bigr) \bigr]^2 + \lambda_1(a -
\widehat
\alpha)^2,
\]
where $\widehat{\alpha}$ is the estimator of $\alpha$ obtained from
the full data. Finally, the approximate
leave-one-curve-out cross-validation score is
\begin{equation}\label{eq:approx_CV}
\widetilde{CV} := \sum_{i=1}^n \sum_{l=1}^{N_i} \sum
_{j=1}^{m_{il}}\ell_{ilj}^{cv}\bigl(\widetilde
a_{il}^{(-il)},\widetilde\theta_i^{(-il)},\widetilde{\bolds{\beta
}}^{(-il)}\bigr).
\end{equation}
\end{appendix}

\section*{Acknowledgments}
The authors would like to thank Professor Wendy Silk of the
Department of Land, Air and Water Resources, University of
California, Davis, for providing the data used in the paper and for
helpful discussions on the scientific aspects of the problem.

\begin{supplement}
\stitle{Supplement to ``Semiparametric modeling of autonomous
nonlinear dynamical systems with application to plant growth''\\}
\slink[doi,text={10.1214/11-AOAS459SUPP}]{10.1214/11-AOAS459SUPP} 
\slink[url]{http://lib.stat.cmu.edu/aoas/459/supplement.pdf}
\sdatatype{.pdf}
\sdescription{The supplementary materials provide additional details
on the computational schemes. It also contains further simulation
studies elucidating the performance of the proposed estimators under
scenarios not covered in the main article.}
\end{supplement}

%

\printaddresses


\begin{thebibliography}{40}

\bibitem[\protect\citeauthoryear{Basu et~al.}{1998}]{Basu2007}
%
\begin{barticle}[auto:STB|2011-03-03|12:04:44]
\bauthor{\bsnm{Basu},~\bfnm{P.}\binits{P.}},
\bauthor{\bsnm{Pal},~\bfnm{A.}\binits{A.}},
\bauthor{\bsnm{Lynch},~\bfnm{J.~P.}\binits{J.~P.}} \AND
\bauthor{\bsnm{Brown},~\bfnm{K.~M.}\binits{K.~M.}}
(\byear{1998}).
\btitle{A novel image-analysis technique for kinematic study of growth and
curvature}.
\bjournal{Plant Physiology}
\bvolume{145}
\bpages{305--316}.
\end{barticle}
%
\endbibitem

\bibitem[\protect\citeauthoryear{Brunel}{2008}]{Brunel2008}
%
\begin{barticle}[mr]
\bauthor{\bsnm{Brunel},~\bfnm{Nicolas J-B.}\binits{N.~J.-B.}}
(\byear{2008}).
\btitle{Parameter estimation of {ODE}'s via nonparametric estimators}.
\bjournal{Electron. J. Statist.}
\bvolume{2}
\bpages{1242--1267}.
\bid{doi={10.1214/07-EJS132}, issn={1935-7524}, mr={2471285}}
\end{barticle}
%
\endbibitem

\bibitem[\protect\citeauthoryear{Burman}{1990}]{Burman1990}
%
\begin{barticle}[mr]
\bauthor{\bsnm{Burman},~\bfnm{Prabir}\binits{P.}}
(\byear{1990}).
\btitle{Estimation of generalized additive models}.
\bjournal{J. Multivariate Anal.}
\bvolume{32}
\bpages{230--255}.
\bid{doi={10.1016/0047-259X(90)90083-T}, issn={0047-259X}, mr={1046767}}
\end{barticle}
%
\endbibitem

\bibitem[\protect\citeauthoryear{Cao, Fussmann and Ramsay}{2008}]{Cao2008}
%
\begin{barticle}[mr]
\bauthor{\bsnm{Cao},~\bfnm{Jiguo}\binits{J.}},
\bauthor{\bsnm{Fussmann},~\bfnm{Gregor~F.}\binits{G.~F.}} \AND
\bauthor{\bsnm{Ramsay},~\bfnm{James~O.}\binits{J.~O.}}
(\byear{2008}).
\btitle{Estimating a predator-prey dynamical model with the parameter cascades
method}.
\bjournal{Biometrics}
\bvolume{64}
\bpages{959--967}.
\bid{doi={10.1111/j.1541-0420.2007.00942.x}, issn={0006-341X}, mr={2526648}}
\end{barticle}
%
\endbibitem

\bibitem[\protect\citeauthoryear{Chen and Wu}{2008a}]{ChenWu2008b}
%
\begin{barticle}[mr]
\bauthor{\bsnm{Chen},~\bfnm{Jianwei}\binits{J.}} \AND
\bauthor{\bsnm{Wu},~\bfnm{Hulin}\binits{H.}}
(\byear{2008}a).
\btitle{Efficient local estimation for time-varying coefficients in
deterministic dynamic models with applications to {HIV}-1 dynamics}.
\bjournal{J. Amer. Statist. Assoc.}
\bvolume{103}
\bpages{369--384}.
\bid{doi={10.1198/016214507000001382}, issn={0162-1459}, mr={2420240}}
\end{barticle}
%
\endbibitem

\bibitem[\protect\citeauthoryear{Chen and Wu}{2008b}]{ChenWu2008a}
%
\begin{barticle}[mr]
\bauthor{\bsnm{Chen},~\bfnm{Jianwei}\binits{J.}} \AND
\bauthor{\bsnm{Wu},~\bfnm{Hulin}\binits{H.}}
(\byear{2008}b).
\btitle{Estimation of time-varying parameters in deterministic dynamic models}.
\bjournal{Statist. Sinica}
\bvolume{18}
\bpages{987--1006}.
\bid{issn={1017-0405}, mr={2440076}}
\end{barticle}
%
\endbibitem

\bibitem[\protect\citeauthoryear{Diggle et~al.}{2002}]{Diggle2002}
%
\begin{bbook}[mr]
\bauthor{\bsnm{Diggle},~\bfnm{Peter~J.}\binits{P.~J.}},
\bauthor{\bsnm{Heagerty},~\bfnm{Patrick~J.}\binits{P.~J.}},
\bauthor{\bsnm{Liang},~\bfnm{Kung-Yee}\binits{K.-Y.}} \AND
\bauthor{\bsnm{Zeger},~\bfnm{Scott~L.}\binits{S.~L.}}
(\byear{2002}).
\btitle{Analysis of Longitudinal Data},
\bedition{2nd} ed.
\bseries{Oxford Statistical Science Series}
\bvolume{25}.
\bpublisher{Oxford Univ. Press}, \baddress{Oxford}.
\bid{mr={2049007}}
\end{bbook}
%
\endbibitem

\bibitem[\protect\citeauthoryear{Fraser, Silk and Rost}{1990}]{Fraser1990}
%
\begin{barticle}[auto:STB|2011-03-03|12:04:44]
\bauthor{\bsnm{Fraser},~\bfnm{T.~K.}\binits{T.~K.}},
\bauthor{\bsnm{Silk},~\bfnm{W.~K.}\binits{W.~K.}} \AND
\bauthor{\bsnm{Rost},~\bfnm{T.~L.}\binits{T.~L.}}
(\byear{1990}).
\btitle{Effects of low water potential on cortical cell length in growing
regions of maize roots}.
\bjournal{Plant Physiology}
\bvolume{93}
\bpages{648--651}.
\end{barticle}
%
\endbibitem

\bibitem[\protect\citeauthoryear{Gu}{2002}]{Gu2002}
%
\begin{bbook}[mr]
\bauthor{\bsnm{Gu},~\bfnm{Chong}\binits{C.}}
(\byear{2002}).
\btitle{Smoothing Spline {ANOVA} Models}.
\bpublisher{Springer}, \baddress{New York}.
\bid{mr={1876599}}
\end{bbook}
%
\endbibitem

\bibitem[\protect\citeauthoryear{Guedj, Thi{\'e}baut and
Commenges}{2007}]{Guedj2007}
%
\begin{barticle}[mr]
\bauthor{\bsnm{Guedj},~\bfnm{J.}\binits{J.}},
\bauthor{\bsnm{Thi{\'e}baut},~\bfnm{R.}\binits{R.}} \AND
\bauthor{\bsnm{Commenges},~\bfnm{D.}\binits{D.}}
(\byear{2007}).
\btitle{Maximum likelihood estimation in dynamical models of {HIV}}.
\bjournal{Biometrics}
\bvolume{63}
\bpages{1198--1206, 1314}.
\bid{issn={0006-341X}, mr={2414598}}
\end{barticle}
%
\endbibitem


\bibitem[\protect\citeauthoryear{Kaslow et~al.}{1987}]{Kaslow1987}
%
\begin{barticle}[pbm]
\bauthor{\bsnm{Kaslow},~\bfnm{R.~A.}\binits{R.~A.}},
\bauthor{\bsnm{Ostrow},~\bfnm{D.~G.}\binits{D.~G.}},
\bauthor{\bsnm{Detels},~\bfnm{R.}\binits{R.}},
\bauthor{\bsnm{Phair},~\bfnm{J.~P.}\binits{J.~P.}},
\bauthor{\bsnm{Polk},~\bfnm{B.~F.}\binits{B.~F.}} \AND
\bauthor{\bsnm{Rinaldo},~\bsuffix{Jr.}, \bfnm{C.~R.}\binits{C.~R.}}
(\byear{1987}).
\btitle{The Multicenter AIDS Cohort Study: Rationale, organization, and
selected characteristics of the participants}.
\bjournal{Am. J. Epidemiol.}
\bvolume{126}
\bpages{310--318}.
\bid{issn={0002-9262}, pmid={3300281}}
\end{barticle}
%
\endbibitem

\bibitem[\protect\citeauthoryear{Ke and Wang}{2001}]{Ke2001}
%
\begin{barticle}[mr]
\bauthor{\bsnm{Ke},~\bfnm{Chunlei}\binits{C.}} \AND
\bauthor{\bsnm{Wang},~\bfnm{Yuedong}\binits{Y.}}
(\byear{2001}).
\btitle{Semiparametric nonlinear mixed-effects models and their applications}.
\bjournal{J. Amer. Statist. Assoc.}
\bvolume{96}
\bpages{1272--1298}.
\bid{doi={10.1198/016214501753381913}, issn={0162-1459}, mr={1946577}}
\bptnote{check related}
\end{barticle}
%
\endbibitem

\bibitem[\protect\citeauthoryear{Lee, Nelder and Pawitan}{2006}]{Lee2006}
%
\begin{bbook}[mr]
\bauthor{\bsnm{Lee},~\bfnm{Youngjo}\binits{Y.}},
\bauthor{\bsnm{Nelder},~\bfnm{John~A.}\binits{J.~A.}} \AND
\bauthor{\bsnm{Pawitan},~\bfnm{Yudi}\binits{Y.}}
(\byear{2006}).
\btitle{Generalized Linear Models with Random Effects: Unified
Analysis via $H$-Likelihood}.
\bseries{Monographs on Statistics and Applied Probability}
\bvolume{106}.
\bpublisher{Chapman \& Hall/CRC}, \baddress{Boca Raton, FL}.
\bid{doi={10.1201/9781420011340}, mr={2259540}}
\end{bbook}
%
\endbibitem

\bibitem[\protect\citeauthoryear{Li et~al.}{2002}]{Li2002}
%
\begin{barticle}[mr]
\bauthor{\bsnm{Li},~\bfnm{Lang}\binits{L.}},
\bauthor{\bsnm{Brown},~\bfnm{Morton~B.}\binits{M.~B.}},
\bauthor{\bsnm{Lee},~\bfnm{Kyung-Hoon}\binits{K.-H.}} \AND
\bauthor{\bsnm{Gupta},~\bfnm{Suneel}\binits{S.}}
(\byear{2002}).
\btitle{Estimation and inference for a spline-enhanced population
pharmacokinetic model}.
\bjournal{Biometrics}
\bvolume{58}
\bpages{601--611}.
\bid{doi={10.1111/j.0006-341X.2002.00601.x}, issn={0006-341X}, mr={1933534}}
\end{barticle}
%
\endbibitem

\bibitem[\protect\citeauthoryear{Ljung and Glad}{1994}]{LjungGlad1994}
%
\begin{bbook}[auto:STB|2011-03-03|12:04:44]
\bauthor{\bsnm{Ljung},~\bfnm{L.}\binits{L.}} \AND
\bauthor{\bsnm{Glad},~\bfnm{T.}\binits{T.}}
(\byear{1994}).
\btitle{Modeling of Dynamic Systems}.
\bpublisher{Prentice Hall}, \baddress{Englewood Cliffs, NJ}.
\end{bbook}
%
\endbibitem

\bibitem[\protect\citeauthoryear{Miao et~al.}{2009}]{Miao2009}
%
\begin{barticle}[pbm]
\bauthor{\bsnm{Miao},~\bfnm{Hongyu}\binits{H.}},
\bauthor{\bsnm{Dykes},~\bfnm{Carrie}\binits{C.}},
\bauthor{\bsnm{Demeter},~\bfnm{Lisa~M.}\binits{L.~M.}} \AND
\bauthor{\bsnm{Wu},~\bfnm{Hulin}\binits{H.}}
(\byear{2009}).
\btitle{Differential equation modeling of HIV viral fitness
experiments: Model
identification, model selection, and multimodel inference}.
\bjournal{Biometrics}
\bvolume{65}
\bpages{292--300}.
\bid{doi={10.1111/j.1541-0420.2008.01059.x}, issn={1541-0420},
mid={NIHMS170494}, pii={BIOM1059}, pmcid={2838508}, pmid={18510656}}
\end{barticle}
%
\endbibitem

\bibitem[\protect\citeauthoryear{Nocedal and
Wright}{2006}]{NocedalWright2006}
%
\begin{bbook}[mr]
\bauthor{\bsnm{Nocedal},~\bfnm{Jorge}\binits{J.}} \AND
\bauthor{\bsnm{Wright},~\bfnm{Stephen~J.}\binits{S.~J.}}
(\byear{2006}).
\btitle{Numerical Optimization},
\bedition{2nd} ed.
\bpublisher{Springer}, \baddress{New York}.
\bid{mr={2244940}}
\end{bbook}
%
\endbibitem

\bibitem[\protect\citeauthoryear{Nowak and May}{2000}]{NowakMay2000}
%
\begin{bbook}[mr]
\bauthor{\bsnm{Nowak},~\bfnm{Martin~A.}\binits{M.~A.}} \AND
\bauthor{\bsnm{May},~\bfnm{Robert~M.}\binits{R.~M.}}
(\byear{2000}).
\btitle{Virus Dynamics: Mathematical Principles of Immunology and Virology}.
\bpublisher{Oxford Univ. Press}, \baddress{Oxford}.
\bid{mr={2009143}}
\end{bbook}
%
\endbibitem


\bibitem[\protect\citeauthoryear{Paul, Peng and Burman}{2009}]{Paul2009}
%
\begin{bmisc}[auto:STB|2011-03-03|12:04:44]
\bauthor{\bsnm{Paul},~\bfnm{D.}\binits{D.}},
\bauthor{\bsnm{Peng},~\bfnm{J.}\binits{J.}} \AND
\bauthor{\bsnm{Burman},~\bfnm{P.}\binits{P.}}
(\byear{2009}).
\bhowpublished{Semiparametric modeling of autonomous nonlinear dynamical systems
with applications.
Technical report.
Available at
\url{http://arxiv.org/PS\_cache/arxiv/pdf/0906/0906.3501v1.pdf}.}
\end{bmisc}
%
\endbibitem

\bibitem[\protect\citeauthoryear{Paul, Peng and Burman}{2011}]{Paul-supp2011}
%
\begin{bmisc}[auto:STB|2011-03-03|12:04:44]
\bauthor{\bsnm{Paul},~\bfnm{D.}\binits{D.}},
\bauthor{\bsnm{Peng},~\bfnm{J.}\binits{J.}} \AND
\bauthor{\bsnm{Burman},~\bfnm{P.}\binits{P.}}
(\byear{2011}).
\bhowpublished{Supplement to ``Semiparametric modeling of autonomous nonlinear
dynamical systems with application to plant growth.''
\href{http://dx.doi.org/10.1214/11-AOAS459SUPP}{DOI:10.1214/11-AOAS459SUPP}}.
\end{bmisc}
%
\endbibitem

\bibitem[\protect\citeauthoryear{Peng and Paul}{2009}]{PengPaul2009}
%
\begin{barticle}[mr]
\bauthor{\bsnm{Peng},~\bfnm{Jie}\binits{J.}} \AND
\bauthor{\bsnm{Paul},~\bfnm{Debashis}\binits{D.}}
(\byear{2009}).
\btitle{A geometric approach to maximum likelihood estimation of the functional
principal components from sparse longitudinal data}.
\bjournal{J. Comput. Graph. Statist.}
\bvolume{18}
\bpages{995--1015}.
\bid{doi={10.1198/jcgs.2009.08011}, issn={1061-8600}, mr={2598035}}
\end{barticle}
%
\endbibitem


\bibitem[\protect\citeauthoryear{Perthame}{2007}]{Perthame2007}
%
\begin{bbook}[mr]
\bauthor{\bsnm{Perthame},~\bfnm{Beno{\^{\i}}t}\binits{B.}}
(\byear{2007}).
\btitle{Transport Equations in Biology}.
\bpublisher{Birkh\"auser}, \baddress{Basel}.
\bid{mr={2270822}}
\end{bbook}
%
\endbibitem


\bibitem[\protect\citeauthoryear{Poyton et~al.}{2006}]{Poyton2006}
%
\begin{barticle}[auto:STB|2011-03-03|12:04:44]
\bauthor{\bsnm{Poyton},~\bfnm{A.~A.}\binits{A.~A.}},
\bauthor{\bsnm{Varziri},~\bfnm{M.~S.}\binits{M.~S.}},
\bauthor{\bsnm{McAuley},~\bfnm{K.~B.}\binits{K.~B.}},
\bauthor{\bsnm{McLellan},~\bfnm{P.~J.}\binits{P.~J.}} \AND
\bauthor{\bsnm{Ramsay},~\bfnm{J.~O.}\binits{J.~O.}}
(\byear{2006}).
\btitle{Parameter estimation in continuous dynamic models using principal
differential analysis}.
\bjournal{Computers \& Chemical Engineering}
\bvolume{30}
\bpages{698--708}.
\end{barticle}
%
\endbibitem

\bibitem[\protect\citeauthoryear{Ramsay and
Silverman}{2002}]{RamsaySilverman2002}
%
\begin{bbook}[mr]
\bauthor{\bsnm{Ramsay},~\bfnm{J.~O.}\binits{J.~O.}} \AND
\bauthor{\bsnm{Silverman},~\bfnm{B.~W.}\binits{B.~W.}}
(\byear{2002}).
\btitle{Applied Functional Data Analysis}.
\bpublisher{Springer}, \baddress{New York}.
\bid{doi={10.1007/b98886}, mr={1910407}}
\end{bbook}
%
\endbibitem

\bibitem[\protect\citeauthoryear{Ramsay and
Silverman}{2005}]{RamsaySilverman2005}
%
\begin{bbook}[mr]
\bauthor{\bsnm{Ramsay},~\bfnm{J.~O.}\binits{J.~O.}} \AND
\bauthor{\bsnm{Silverman},~\bfnm{B.~W.}\binits{B.~W.}}
(\byear{2005}).
\btitle{Functional Data Analysis},
\bedition{2nd} ed.
\bpublisher{Springer}, \baddress{New York}.
\bid{mr={2168993}}
\end{bbook}
%
\endbibitem

\bibitem[\protect\citeauthoryear{Ramsay et~al.}{2007}]{Ramsay2007}
%
\begin{barticle}[mr]
\bauthor{\bsnm{Ramsay},~\bfnm{J.~O.}\binits{J.~O.}},
\bauthor{\bsnm{Hooker},~\bfnm{G.}\binits{G.}},
\bauthor{\bsnm{Campbell},~\bfnm{D.}\binits{D.}} \AND
\bauthor{\bsnm{Cao},~\bfnm{J.}\binits{J.}}
(\byear{2007}).
\btitle{Parameter estimation for differential equations: A generalized
smoothing approach}.
\bjournal{J. R. Stat. Soc. Ser. B Stat. Methodol.}
\bvolume{69}
\bpages{741--796}.
\bid{doi={10.1111/j.1467-9868.2007.00610.x}, issn={1369-7412}, mr={2368570}}
\bptnote{check related}
\end{barticle}
%
\endbibitem

\bibitem[\protect\citeauthoryear{Sacks, Silk and Burman}{1997}]{Sacks1997}
%
\begin{barticle}[pbm]
\bauthor{\bsnm{Sacks},~\bfnm{M.~M.}\binits{M.~M.}},
\bauthor{\bsnm{Silk},~\bfnm{W.~K.}\binits{W.~K.}} \AND
\bauthor{\bsnm{Burman},~\bfnm{P.}\binits{P.}}
(\byear{1997}).
\btitle{Effect of water stress on cortical cell division rates within the
apical meristem of primary roots of maize}.
\bjournal{Plant Physiol.}
\bvolume{114}
\bpages{519--527}.
\bid{issn={1532-2548}, pii={114/2/519}, pmcid={158332}, pmid={12223725}}
\end{barticle}
%
\endbibitem

\bibitem[\protect\citeauthoryear{Schurr, Walter and
Rascher}{2006}]{Schurr2006}
%
\begin{barticle}[auto:STB|2011-03-03|12:04:44]
\bauthor{\bsnm{Schurr},~\bfnm{U.}\binits{U.}},
\bauthor{\bsnm{Walter},~\bfnm{A.}\binits{A.}} \AND
\bauthor{\bsnm{Rascher},~\bfnm{U.}\binits{U.}}
(\byear{2006}).
\btitle{Functional dynamics of plant growth and photosynthesis---from
steady-state to dynamics---from homogeneity to heterogeneity}.
\bjournal{Plant, Cell and Environment}
\bvolume{29}
\bpages{340--352}.
\end{barticle}
%
\endbibitem

\bibitem[\protect\citeauthoryear{Silk}{1994}]{Silk1994}
%
\begin{barticle}[auto:STB|2011-03-03|12:04:44]
\bauthor{\bsnm{Silk},~\bfnm{W.~K.}\binits{W.~K.}}
(\byear{1994}).
\btitle{Kinametics and dynamics of primary growth}.
\bjournal{Biomimectics}
\bvolume{2}
\bpages{199--213}.
\end{barticle}
%
\endbibitem

\bibitem[\protect\citeauthoryear{Silk and Erickson}{1979}]{Silk1979}
%
\begin{barticle}[auto:STB|2011-03-03|12:04:44]
\bauthor{\bsnm{Silk},~\bfnm{W.~K.}\binits{W.~K.}} \AND
\bauthor{\bsnm{Erickson},~\bfnm{R.~O.}\binits{R.~O.}}
(\byear{1979}).
\btitle{Kinametics of plant growth}.
\bjournal{J. Theoret. Biol.}
\bvolume{76}
\bpages{481--501}.
\end{barticle}
%
\endbibitem

\bibitem[\protect\citeauthoryear{Strogatz}{2001}]{Strogatz2001}
%
\begin{bmisc}[auto:STB|2011-03-03|12:04:44]
\bauthor{\bsnm{Strogatz},~\bfnm{S.~H.}\binits{S.~H.}}
(\byear{2001}).
\bhowpublished{\textit{Nonlinear Dynamics and Chaos: With
Applications to
Physics, Biology, Chemistry and Engineering}. Perseus Books Group, New York}.
\end{bmisc}
%
\endbibitem

\bibitem[\protect\citeauthoryear{Tenenbaum and Pollard}{1985}]{Tenenbaum1985}
%
\begin{bmisc}[auto:STB|2011-03-03|12:04:44]
\bauthor{\bsnm{Tenenbaum},~\bfnm{M.}\binits{M.}} \AND
\bauthor{\bsnm{Pollard},~\bfnm{H.}\binits{H.}}
(\byear{1985}).
\bhowpublished{\textit{Ordinary Differential Equations}. Dover}.
\end{bmisc}
%
\endbibitem

\bibitem[\protect\citeauthoryear{Varah}{1982}]{Varah1982}
%
\begin{barticle}[mr]
\bauthor{\bsnm{Varah},~\bfnm{J.~M.}\binits{J.~M.}}
(\byear{1982}).
\btitle{A spline least squares method for numerical parameter
estimation in
differential equations}.
\bjournal{SIAM J. Sci. Statist. Comput.}
\bvolume{3}
\bpages{28--46}.
\bid{doi={10.1137/0903003}, issn={0196-5204}, mr={0651865}}
\end{barticle}
%
\endbibitem

\bibitem[\protect\citeauthoryear{Walter et~al.}{2002}]{Walter2002}
%
\begin{barticle}[auto:STB|2011-03-03|12:04:44]
\bauthor{\bsnm{Walter},~\bfnm{A.}\binits{A.}},
\bauthor{\bsnm{Spies},~\bfnm{H.}\binits{H.}},
\bauthor{\bsnm{Terjung},~\bfnm{S.}\binits{S.}},
\bauthor{\bsnm{K{\"u}sters},~\bfnm{R.}\binits{R.}},
\bauthor{\bsnm{Kirchgebner},~\bfnm{N.}\binits{N.}} \AND
\bauthor{\bsnm{Schurr},~\bfnm{U.}\binits{U.}}
(\byear{2002}).
\btitle{Spatio-temporal dynamics of expansion growth in roots: Automatic
quantification of diurnal course and temperature response by digital image
sequence processing}.
\bjournal{J. Experimental Botany}
\bvolume{53}
\bpages{689--698}.
\end{barticle}
%
\endbibitem

\bibitem[\protect\citeauthoryear{Wu and Ding}{1999}]{WuDing1999}
%
\begin{barticle}[auto:STB|2011-03-03|12:04:44]
\bauthor{\bsnm{Wu},~\bfnm{H.}\binits{H.}} \AND
\bauthor{\bsnm{Ding},~\bfnm{A.}\binits{A.}}
(\byear{1999}).
\btitle{Population HIV-1 dynamics in vivo: Applicable models and inferential
tools for virological data from AIDS clinical trials}.
\bjournal{Biometrics}
\bvolume{55}
\bpages{410--418}.
\end{barticle}
%
\endbibitem

\bibitem[\protect\citeauthoryear{Wu, Ding and DeGruttola}{1998}]{Wu1998}
%
\begin{barticle}[auto:STB|2011-03-03|12:04:44]
\bauthor{\bsnm{Wu},~\bfnm{H.}\binits{H.}},
\bauthor{\bsnm{Ding},~\bfnm{A.}\binits{A.}} \AND
\bauthor{\bsnm{DeGruttola},~\bfnm{V.}\binits{V.}}
(\byear{1998}).
\btitle{Estimation of HIV dynamic parameters}.
\bjournal{Stat. Med.}
\bvolume{17}
\bpages{2463--2485}.
\end{barticle}
%
\endbibitem

\bibitem[\protect\citeauthoryear{Zhu and Wu}{2007}]{ZhuWu2007}
%
\begin{barticle}[mr]
\bauthor{\bsnm{Zhu},~\bfnm{Haihong}\binits{H.}} \AND
\bauthor{\bsnm{Wu},~\bfnm{Hulin}\binits{H.}}
(\byear{2007}).
\btitle{Estimation of smooth time-varying parameters in state space models}.
\bjournal{J. Comput. Graph. Statist.}
\bvolume{16}
\bpages{813--832}.
\bid{doi={10.1198/106186007X255991}, issn={1061-8600}, mr={2412484}}
\end{barticle}
%
\endbibitem

\end{thebibliography}
\end{document}